\documentclass[aps,prl,superscriptaddress,reprint,showpacs,floatfix]{revtex4-1}
\usepackage{graphicx}
\usepackage{dcolumn}
\usepackage{bm}
\usepackage{xcolor}
\usepackage{relsize}
\usepackage{amsmath}
\usepackage{amsfonts}
\usepackage{mathtools}
\usepackage{braket}
\usepackage{gensymb}
\usepackage{url}
\usepackage{hyperref}
\usepackage{footmisc}
\usepackage[utf8]{inputenc}
\begin{document}

\preprint{APS/123-QED}

\title{Effect of photonic spin-orbit coupling on the topological edge modes of a Su-Schrieffer-Heeger chain}

\author{C. E. Whittaker}
\email{charles.whittaker@sheffield.ac.uk}
\affiliation{Department of Physics and Astronomy, University of Sheffield, Sheffield S3 7RH, United Kingdom}%
\author{E. Cancellieri}
\email{e.cancellieri@lancaster.ac.uk}
\affiliation{Department of Physics and Astronomy, University of Sheffield, Sheffield S3 7RH, United Kingdom}%
\affiliation{Department of Physics, Lancaster University, Lancaster LA1 4YB, UK}%
\author{P. M. Walker}
\affiliation{Department of Physics and Astronomy, University of Sheffield, Sheffield S3 7RH, United Kingdom}%
\author{B. Royall}
\affiliation{Department of Physics and Astronomy, University of Sheffield, Sheffield S3 7RH, United Kingdom}%
\author{L. E. Tapia Rodriguez}
\affiliation{Department of Physics and Astronomy, University of Sheffield, Sheffield S3 7RH, United Kingdom}%
\author{E. Clarke}
\affiliation{EPSRC National Epitaxy Facility, University of Sheffield, Sheffield S1 3JD, United Kingdom}
\author{D. M. Whittaker}
\affiliation{Department of Physics and Astronomy, University of Sheffield, Sheffield S3 7RH, United Kingdom}%
\author{H. Schomerus}
\affiliation{Department of Physics, Lancaster University, Lancaster LA1 4YB, UK}%
\author{M. S. Skolnick}
\affiliation{Department of Physics and Astronomy, University of Sheffield, Sheffield S3 7RH, United Kingdom}
\author{D. N. Krizhanovskii}
\affiliation{Department of Physics and Astronomy, University of Sheffield, Sheffield S3 7RH, United Kingdom}

\date{\today}

\begin{abstract}
We study the effect of photonic spin-orbit coupling (SOC) in micropillar lattices on the topological edge states of a one-dimensional chain with a zigzag geometry, corresponding to the Su-Schrieffer-Heeger model equipped with an additional internal degree of freedom. The system combines the strong hopping anisotropy of the $p$-type pillar modes with the large TE-TM splitting in Bragg microcavities. By resolving the photoluminescence emission in energy and polarization we probe the effects of the resulting SOC on the spatial and spectral properties of the edge modes. We find that the edge modes feature a fine structure of states that penetrate by different amounts into the bulk of the chain, depending on the strength of the SOC terms present, thereby opening a route to manipulation of the topological states in the system. 

\end{abstract}

\maketitle

Concepts of band structure topology from solid state physics now play a prominent role in photonics research. Inspired by discoveries in condensed matter systems, topologically insulating and quantum Hall type phases have been realized in analogous photonic contexts using gyromagnetic photonic crystals \cite{Wang2009,PhysRevLett.106.093903}, coupled ring resonator arrays \cite{Hafezi2013,PhysRevLett.113.087403,Mittal2016} and metamaterials \cite{Chen2014,Slobozhanyuk2016} to engineer topological lattice Hamiltonians. In photonic platforms, additional functionalities may be provided by the presence of gain and loss \cite{Poli2015,PhysRevLett.115.040402}, optical nonlinearities \cite{PhysRevLett.117.143901,1367-2630-19-9-095002} and coupling with quantum emitters \cite{Barik666}. Furthermore, the transverse electric (TE) and transverse magnetic (TM) modes of photonic structures, which are typically split in energy, introduce a pseudospin into the system \cite{Khanikaev2012}. The splitting arises due to a $k$-dependent effective magnetic field acting on the polarization of photons \cite{SOLNYSHKOV2016920}. In analogy with electrons in Dresselhaus or Rashba fields, this phenomenon can be described as a photonic spin-orbit coupling (SOC), which may be enhanced in layered or laterally modulated wavelength-scale structures and used to engineer artificial gauge fields \cite{PhysRevLett.114.026803,AIDELSBURGER2018}. 

In Bragg-mirror micropillar arrays, splitting between TE and TM linearly polarized modes is generally sizable meaning photonic SOC is pronounced. It arises due to the fact that for different polarizations of the cavity field there are inequivalent boundary conditions at the layer interfaces in the vertical direction and at the pillar sidewalls in the lateral direction. 
It plays a key role in several recent proposals to engineer topological protection in polariton systems \cite{PhysRevLett.114.116401,Gulevich2017,PhysRevB.97.081103} in addition to emulating spin-dependent phenomena from solid state systems \cite{PhysRevB.93.085404}. Experimentally it has been explored in a hexagonal ring of coupled micropillars whose eigenmodes are spin (polarization) vortices \cite{PhysRevX.5.011034} and a Lieb lattice where polarization textures of flat-band modes were observed \cite{PhysRevLett.120.097401,doi:10.1063/1.4995385}. However, in the case of topological edge modes these photonic SOC effects revealed by the polarization degree of freedom (DOF) remain unexplored in experimental works. 

The Su-Schrieffer-Heeger (SSH) model represents one of the simplest possible systems exhibiting topological edge modes \cite{PhysRevLett.42.1698}, offering a convenient starting point to explore the polarization DOF in the context of topological band structures. It comprises a one-dimensional (1D) dimerized chain with a two-site unit cell, with alternating hopping energies between sites (within and between dimers), analogous to polymer chains where the Peierls instability makes dimerization energetically favourable. In photonic systems, SSH models have been realized in diverse platforms such as coupled waveguides \cite{Malkova:09,LPOR:LPOR201400462}, plasmonic nanodisks \cite{C5NR00231A}, and both passive \cite{Weimann2016} and active \cite{Zhao2018,PhysRevLett.120.113901} SSH-like arrays with additional gain and loss distributions. In GaAs-based micropillar arrays, a variant of the SSH chain which directly uses the native photonic SOC to engineer a staggered hopping of $s$-type pillar modes has been proposed \cite{PhysRevLett.116.046402}. In practice, however, the stringent requirements on both the mode linewidth and magnitude of polarization splitting render the realization of such a model challenging. 

A recent experimental work \cite{St-Jean2017} implemented an \em{orbital}\rm{} version of the SSH model using the strong
staggered hopping potential experienced by the doubly-degenerate first excited pillar modes, $p_x$ and $p_y$. The spatial mode symmetries and geometrical  configuration of the chain 
combine to induce alternating strong and weak bonds between sites. The magnitude of this tunneling anisotropy is sufficient
to open a large gap (many times larger than the linewidth) containing exponentially-localized edge states. The fact that there are two $p$-type modes means that the system actually
constitutes two copies of the SSH model, which are in topologically inequivalent phases, such that edge states
can be observed in both subspaces depending  on the geometry at the ends of the chain. Compared to the case of $s$ modes, the influence of photonic SOC is expected to be even richer when dealing with $p$ modes, due to the possibility of strong polarization splittings in both the on-site and tunneling energies of modes \cite{PhysRevX.5.011034,PhysRevLett.115.246401} which have different effects on the topological properties of the SSH Hamiltonian \cite{asboth2016short}. The strength of the polarization terms can be varied by the layer structure of the Bragg mirrors, making the SOC a flexible tool which, until now, has not been studied in relation to topology. In this Letter, we consider a photonic zigzag chain where both the confinement and tunneling energies of $p$ modes depend strongly on the polarization. We show how in this case the twofold SSH Hamiltonian splits into a novel fourfold variant with significant differences between the two pseudospins (polarizations) as a result of large SOC, which as we demonstrate, can be probed in photoluminescence (PL) experiments by the spectral and spatial properties of the edge modes. We also discuss the general interplay between different polarization effects and how the strength of different perturbations which contribute to the SOC determines the symmetries of the system.  

\begin{figure}[t]
\centering
\includegraphics[width=0.45\textwidth]{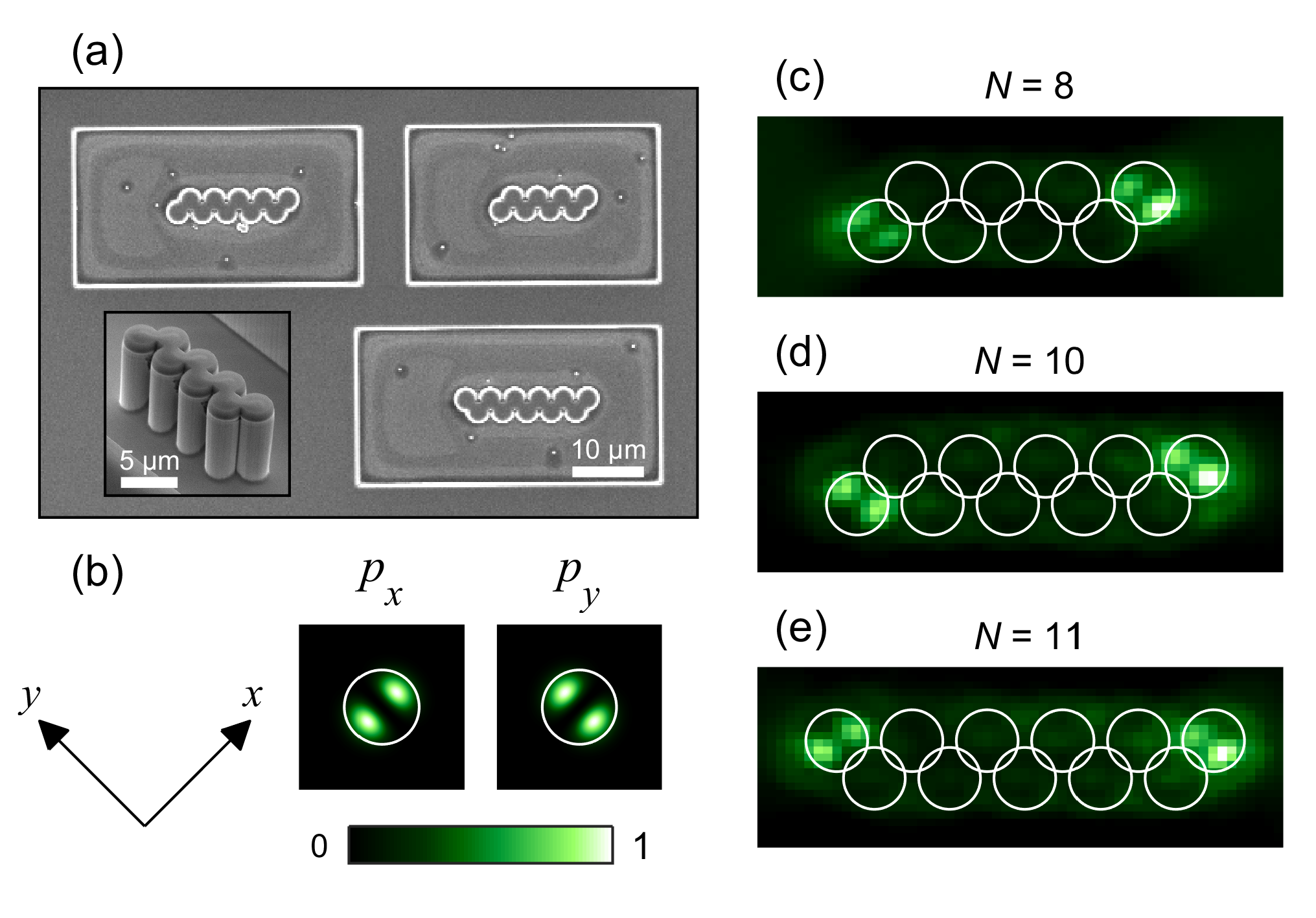}
\caption{(a) Scanning electron microscope image of the photonic zigzag chain structures. The inset shows an angled image of an 8-site chain. (b) Schematic of the $p_x$ and $p_y$ modes of a single micropillar. (c)--(e) Real space images of the topological edge modes for chains with 8, 10 and 11 sites.}
\label{fig1}
\end{figure}

Our sample is a GaAs cavity embedded between GaAs/Al$_{0.85}$Ga$_{0.15}$As distributed Bragg reflectors with 23 (26) top (bottom) pairs, featuring 6 In$_{0.04}$Ga$_{0.96}$As quantum wells. The exciton energy is detuned roughly 20 meV from the cavity mode, and we estimate that the TE-TM splitting has a magnitude on the order of $\beta$ = -0.19 meV$\mu$m$^2$. The resulting large SOC was deliberately designed by using the offset between the Bragg mirror stop band and the cavity mode \cite{PhysRevB.59.5082}, allowing us to enter a qualitatively different regime to that of Ref. \cite{St-Jean2017} and other scalar photonic SSH models. We process our cavity using electron beam lithography and plasma dry etching to create patterned regions with arrays of overlapping micropillars. The pillars have diameters of 3 $\mu$m and centre-to-centre distances of 2.55 $\mu$m. In order to study topological edge modes we consider 1D arrays in a zigzag geometry, with 8, 10 and 11 sites [see Fig. \ref{fig1}(a)]. The number of pillars was chosen in order to minimize variation along the chain length, without being too short to make the ends significant, such that edge states hybridize. The energy-resolved emission from the chains under weak nonresonant excitation shows bands formed from evanescent coupling of both $s$ and $p$ modes of the individual pillars [see Ref. \footnote{See supplementary material at [url] for further details about the experimental measurements, additional experimental data from several microstructures on the same sample, a description of the tight-binding model developed and a discussion of the topological properties of the system. Also includes Refs. \cite{PhysRevLett.81.2582,1367-2630-12-6-065010}} for supplementary information]. A twofold degeneracy comprising $p_x$ and $p_y$ orbitals exists in the latter case, where the subscript refers to the direction in which the bright lobes are oriented [see Fig. \ref{fig1}(b)]. Critically, in arrays of coupled pillars, tunneling between the $p$ modes is strong (weak) when their orientation is aligned (transverse) to the tunneling direction \cite{PhysRevLett.120.097401}, meaning for zigzag chains, where the tunneling direction changes by 90$\degree$ from site to site (alternating between $x$ and $y$), the hopping energies alternate between strong and weak. 

If we neglect the polarization DOF for the moment, our zigzag chains implement a twofold SSH model like the one described in Ref. \cite{St-Jean2017}. 
The manifestation of the topologically non-trivial nature of the chain is spectrally-isolated mid-gap states whose wave functions are strongly confined to the edge pillars and, depending on the number of sites in the chain, can be in one or both of the $p$ orbital subspaces. As can be seen from the real space emission of the edge states in Figs. \ref{fig1}(c) and (d), only the $p_y$ subspace features edge states in our \em{}even\rm{} chains. This is expected since the links connecting the end pillars to the next pillar point along $x$, to which $p_y$ modes are orthogonally oriented, meaning the bond is weak. Conversely, the $p_x$ modes point along $x$ so the bond is strong. Regardless of the choice of unit cell, these two subspaces are topologically inequivalent as determined by the unique difference in the Zak phase \cite{asboth2016short}. In \em{}odd\rm{} chains, edge states are found in both $p_x$ and $p_y$ subspaces, at opposite edges, since the half-integer number of unit cells means there is always a weakly bonded site at one of the edges [see Fig. \ref{fig1}(e)]. Hence, for any number of pillars in a finite chain there are always mid-gap states at both edges, which are found in the same (different) subspaces for even (odd) chains.

\begin{figure*}[t]
\centering
\includegraphics[width=\textwidth]{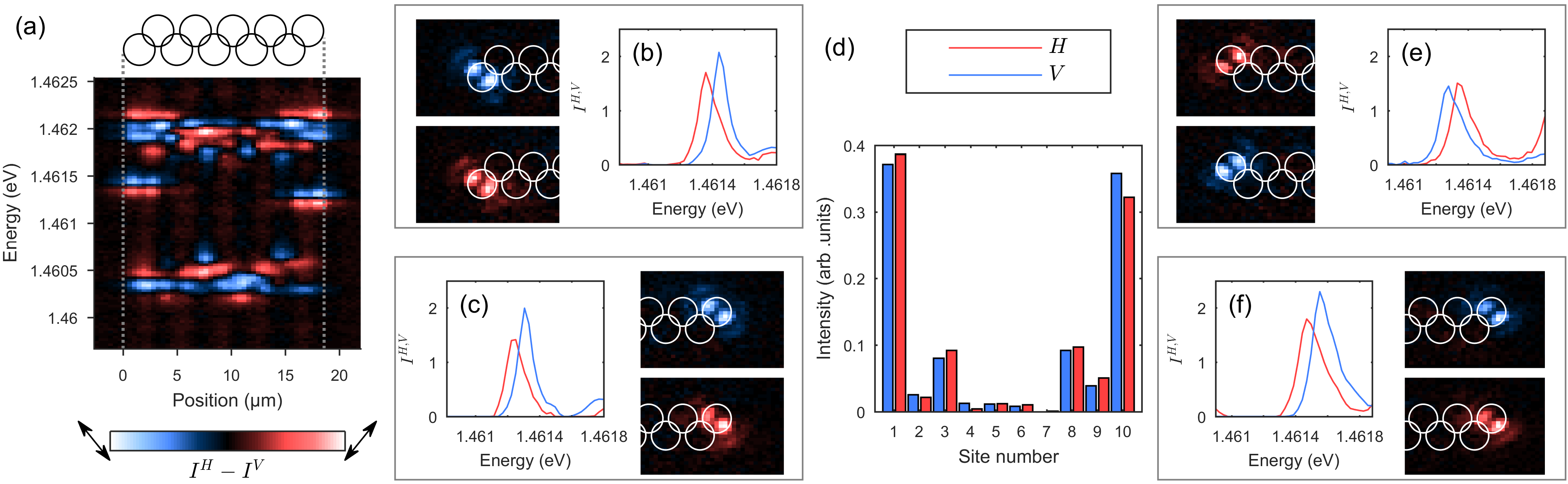}
\caption{(a)--(d) Results for the 10-site chain. (a) Real space spectrum showing the differential polarization intensity $I^H-I^V$. The directions of $H$ and $V$ polarizations are depicted by the arrows. (b) Polarization-resolved intensities $I^{H}$ (red) and $I^{V}$ (blue) of the left edge states, with corresponding real space images on the left. (c) Polarization-resolved spectrum of the right edge states with corresponding real space images on the right. (d) Intensity against site number for the two polarizations across the topological gap. (e),(f) Polarization-resolved spectra for the 11-site chain.}
\label{fig2}
\end{figure*}

Now we will turn our attention to the internal polarization DOF. In this case there are four modes: $p_x$ and $p_y$ in two orthogonal polarizations. When the cavity TE-TM splitting and hence photonic SOC is strong (as quantified by the $\beta$ factor), the $p$-like modes combine into spin-vortices whose energies depend on the sign and size of $\beta$ \cite{PhysRevLett.115.246401}. In our sample, resolving the emission from single pillars in polarization reveals that the $p$ modes have well-defined pseudospin textures and are significantly split in energy. When single pillars are coupled into a dimer, the spectrum of hybridized $p$ modes then shows a marked asymmetry due to the interplay between this on-site polarization splitting and polarization-dependent tunneling. We use the experimental estimates of polarization terms from the single and coupled pillar measurements shown in Ref. \cite{Note1} for our phenomenological tight-binding model later in the text. In the SSH model, the spectral positions of the edge modes are sensitive to on-site perturbations whereas the size of the gap and localization length of the edge modes are affected by perturbations to the tunneling energy. We thus resolve the emission from our zigzag chains in two orthogonal polarizations to see whether we can detect the influence of the polarization DOF on the topological edge modes. 

Fig. \ref{fig2}(a) presents the spectrally-resolved real space emission from the $p$ bands of our 10-site chain, showing the differential polarization intensity corresponding to the difference in emission between in-plane polarizations pointing along $x$ and $y$ respectively, which we define as horizontal ($H$) and vertical ($V$). The significant degree of polarization of the emission (on the order of 15--20$\%$) is immediately evident, demonstrating the large degeneracy lifting created by the combination of polarization effects. Note that there is an energy difference between the left and right edge modes, which probably arises due to a combination of the cavity wedge, etching-induced strain and disorder. For clarity, we henceforth treat the left and right edges of the chain separately, and note that the energy gradient does not affect our subsequent analysis. In Figs. \ref{fig2}(b) and (c) we show the polarization-resolved spectra of the left and right edge states respectively, where two peaks with a splitting on the order of 0.1 meV can be seen in both cases. Alongside these spectra we plot the differential polarization intensity of the real space emission at the energies of the peaks. Since both edge states are found in the $p_y$ subspace, we expect the same sign of polarization splitting at both ends of the chain, which we indeed observe in experiment. In contrast, in odd chains the left and right edge states are orthogonally-oriented with respect to each other, so the sign of polarization splitting is opposite at the two ends [see Figs. \ref{fig2}(e) and (f)]. 

In order to determine a polarization splitting in the hopping energy of modes one may consider the spatial profile of the edge states. In standard SSH theory, they are exponentially localized with a wave function given by $|\Psi_n|^2 \propto (t/t')^{n}$ if $n$ is odd and by $\Psi=0$ if $n$ is even, where $n$ denotes the pillar number counting from 1 and starting at the edge, and $t$ and $t'$ give the tunneling energies within and between unit cells respectively. Since photonic SOC lifts the degeneracy of both $t$ and $t'$ between orthogonal polarizations, a disparity should exist between the spatial wave functions of the orthogonally-polarized topological edge modes in our zigzag chains. In order to see if this is the case in experiment, we estimate the mode intensity (which is proportional to $|\Psi|^2$) against site number from our energy-resolved real space data and then compare between the two polarizations. We show the result for our 10-site chain in Fig. \ref{fig2}(d), which presents the peak intensity on each site for the two polarizations within the gap between lower and upper $p$ bands. As expected theoretically, the edge-state wave function is almost entirely localized on the first and third pillars (counting from either edge) such that sites 1 and 3 (left edge) and 8 and 10 (right edge) have the highest intensities. Since the population on these sites is most significant, we use the values from these sites to quantify the effect of the polarization-dependent tunneling by defining a quantity

\begin{equation}
\xi = \frac{|\Psi_1^{\parallel}|^2/|\Psi_3^{\parallel}|^2}{|\Psi_1^{\perp}|^2/|\Psi_3^{\perp}|^2},
\end{equation}
\noindent
where the subscript denotes the pillar number counted from the edge (left or right) and the superscript denotes parallel ($\parallel$) and perpendicular ($\perp$) polarizations. The value of $\xi$ then gives a quantitative measure of the ratio of the wave function decay between the two polarizations, i.e. SOC in the hoppings. Incorporating the tunneling values and polarization-dependent corrections extracted from our single dimer measurements \cite{Note1} into a conventional SSH model for fixed polarization (see above expression for $\Psi_n$) we can estimate a theoretical figure yielding $\xi^{\rm{theory}} \approx 0.8$. Physically this tells us that the \em{}inverse\rm{} localization length should be shorter for parallel polarization, i.e. the edge state penetrates more into the rest of the chain when its polarization is parallel to the tunneling link. By considering all three of our zigzag chains, we have six experimental values of $\xi$ since each chain has mid-gap states at both edges. In the case of the 11-site chain, $\parallel$ and $\perp$ polarizations are different at the two edges. In all cases $\xi$ is found to lie between 0.7 and 0.9 with an average of $\xi^{\rm{exp.}} = 0.78 \pm 0.07$, which is in good agreement with the ratio of tunneling rates obtained from the spectrum of the dimer.

To obtain a more detailed understanding of the experimental findings, we develop a tight-binding model 
\begin{equation}
H_{\rm{SSH}} = H_{0} + H_{\tau_x} + H_{\tau_y} + H_{\tau_m}
\end{equation}
that systematically accounts for all polarization effects across the full structure. Denoting by ${\hat p}^{H}_{x,n}$, ${\hat p}^{H}_{y,n}$, ${\hat 
p}^{V}_{x,n}$, ${\hat p}^{V}_{y,n}$ the annihilation operators of the $p$ orbitals on pillar $n$ with linear polarization  $H$ (along $x$) and $V$ (along $y$), the Hamiltonian for the chain of isolated pillars is given by
\begin{equation}
\begin{aligned}
H_{0} = \Delta E\sum_{n=1}^{N}[\hat{p}^{H\dagger}_{x,n}\hat{p}^{H}_{x,n}-\hat{p}^{H\dagger}_{y,n}\hat{p}^{H}_{y,n}-\hat{p}^{V\dagger}_{x,n}\hat{p}^{V}_{x,n}+\hat{p}^{V\dagger}_{y,n}\hat{p}^{V}_{y,n}] \\
+\Delta E\sum_{n=1}^{N}[\hat{p}^{H\dagger}_{x,n}\hat{p}^{V}_{y,n}+\hat{p}^{H\dagger}_{y,n}\hat{p}^{V}_{x,n}+\rm{h.c.}], 
\end{aligned}
\end{equation}
\noindent
where $\Delta E$ is the SOC matrix element of a single pillar  and $n$ indicates the pillar number. 
The coupling between neighboring pillars along the $x$ direction is given by
\begin{equation}
\begin{aligned}
H_{\tau_x} = \sum_{i=1}^{N/2}[\tau_{a}^{\parallel}\hat{p}^{H\dagger}_{x,2i-1}\hat{p}^{H}_{x,2i}+\tau_{t}^{\parallel}\hat{p}^{H\dagger}_{y,2i-1}\hat{p}^{H}_{y,2i} \\
+\tau_{a}^{\perp}\hat{p}^{V\dagger}_{x,2i-1}\hat{p}^{V}_{x,2i}+\tau_{t}^{\perp}\hat{p}^{V\dagger}_{y,2i-1}\hat{p}^{V}_{y,2i}],
\end{aligned}
\end{equation}
\noindent
where $\tau_{a(t)}^{\parallel}$ and $\tau_{a(t)}^{\perp}$ describe the coupling of $p$ orbitals whose lobes are aligned ($a$) or transverse ($t$) to the coupling direction, while their  polarization is either parallel ($\parallel$) or perpendicular ($\perp$) to this direction. The coupling term $H_{\tau_y}$ along the $y$ direction is obtained by interchanging $\tau_{a}^{\parallel}$  with $\tau_{t}^{\perp}$ and $\tau_{a}^{\perp}$  with $\tau_{t}^{\parallel}$. 
Finally, the term 
\begin{equation}
H_{\tau_m} = \tau_m\sum_{i=1}^{N/2}[\hat{p}^{H\dagger}_{x,2i-1}\hat{p}^{V}_{y,2i}+\hat{p}^{H\dagger}_{y,2i-1}\hat{p}^{V}_{x,2i}+\rm{h.c.}]
\end{equation}
describes the mixing of $H$-polarized $p_x$ ($p_y$) orbitals with $V$-polarized $p_y$ ($p_x$) orbitals.

The structure of these terms follows from symmetry considerations, while the values of the matrix elements can be estimated in perturbation theory. For this, we represent the $p$ orbitals as the first excited states
$p_x(x,y)=(2/\pi)^{1/2}m\omega xe^{-m\omega (x^2+y^2)/2}$, $p_y(x,y)=(2/\pi)^{1/2}m\omega y e^{-m\omega (x^2+y^2)/2}$ of a two-dimensional harmonic oscillator with potential $U(x,y)=\frac{1}{2}m\omega^2(x^2+y^2)$ and harmonic confinement strength $\omega$ for polaritons of mass $m$, with $\hbar=1$. Centering these parabolic potentials at each pillar determines the barrier shape, for which  the perturbative matrix elements can be evaluated analytically
\cite{Note1}.  The theoretical values can then be matched to the experimental polarization-resolved PL data for a single pillar and dimer, which provides an estimate of $\Delta{E}$ and $\tau_a^{\parallel}$, $\tau_a^{\perp}$, $\tau_t^{\parallel}$, $\tau_t^{\perp}$, $\tau_m$, respectively.

\begin{figure}[t]
\centering
\includegraphics[width=0.5\textwidth]{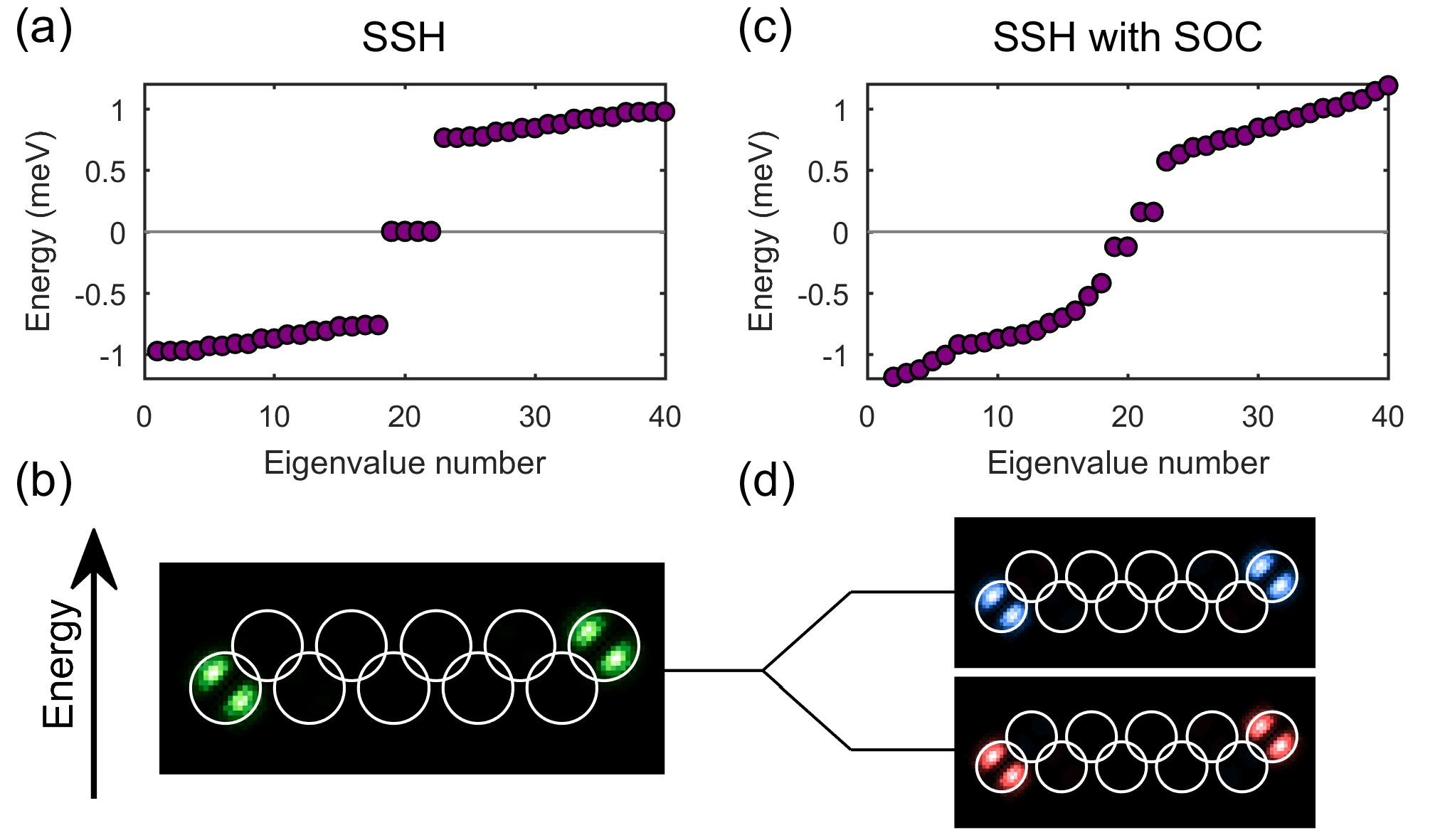}
\caption{Energy spectra and real space images of mid-gap states from the tight-binding model of a 10-site chain without TE-TM splitting (a),(b), and with TE-TM splittings both on-site and in the coupling term (c),(d). In (d), the intensity distribution is resolved in polarization ($V$: top, $H$: bottom).}
\label{zigzag_N10_theory}
\end{figure}

Fig. \ref{zigzag_N10_theory} shows the results obtained from this approach for a zigzag chain with 10 sites.  Panels (a) and (b) show the energies and edge state mode profile for the case without polarization, which corresponds to the case realized in Ref. \cite{St-Jean2017}. Panel (c) shows the energies when all  polarization effects are taken into account. As in the experiments, the edge states are split, with the lower eigenvalue being $H$-polarized while the higher one is $V$-polarized. The differential polarization real space images in panel (d) agree well with the experimental results shown in Figs. \ref{fig2}(a)--(c). From the different localization lengths of the edge states we find $\xi^{\rm{TB}} \approx 0.76$, which is consistent with our earlier estimate $\xi^{\rm{theory}}$ and the experimental value $\xi^{\rm{exp.}}$.

Based on this TB model, we can assess how the
SOC determines the topology of the polarization-resolved
modes of the system. For $\Delta E\ne 0$ but $\tau_m=0$, 
$\tau_a^{\parallel}=\tau_a^{\perp}$, and 
$\tau_t^{\parallel}=\tau_t^{\perp}$ (i.e. the SOC affects the on-site energies but not the couplings),  the system realizes a fourfold SSH model with energy splittings replicating the spin-vortex states of a single pillar. For  $\Delta E=0$ but $\tau_m\ne 0$, $\tau_a^{\parallel}\neq\tau_a^{\perp}$, and $\tau_t^{\parallel}\neq\tau_t^{\perp}$, we again realize four copies, but with polarization-dependent couplings as quantified by $\xi\neq 1$. In our experiments, $\Delta{E}$ is of the same order of magnitude as the linewidth, and much smaller than the band gap,
meaning the topology of the system is only weakly violated. When one further departs from these conditions, the system crosses over to a topologically trivial insulator of the $AI$ symmetry class \cite{Ryu2010,PhysRevB.90.165114,Note1}.

In conclusion, our work suggests that the polarization degree of freedom could be used as a powerful tool to control the topology in a wide range of 1D and 2D lattice systems. Moreover, by probing both the spectral and spatial polarization properties of topological edge states, information about the energy splittings in the pillars and effect on bulk transport can be retrieved. This is also particularly interesting due to the fact that it is possible to control the polarization splitting of $p$ orbitals through the layer structure of the Bragg mirrors \cite{Note1}. Finally, using samples with a less negative cavity-exciton detuning (leading to polaritons with a much larger exciton fraction) will also allow further manipulation of the energy bands through nonlinear renormalization in high-density regimes, the Zeeman effect under application of a magnetic field and via ultrafast Stark control \cite{PhysRevB.94.201301}, making our system a unique test bed to investigate topological phase transitions in exotic lattice Hamiltonians with spin-orbit coupling, interparticle interactions and broken time-reversal symmetry.

The work was supported by EPSRC Grant EP/N031776/1 and ERC Advanced Grant EXCIPOL No.
320570. 

\bibliography{references}   

\begin{thebibliography}{41}%
\makeatletter
\providecommand \@ifxundefined [1]{%
 \@ifx{#1\undefined}
}%
\providecommand \@ifnum [1]{%
 \ifnum #1\expandafter \@firstoftwo
 \else \expandafter \@secondoftwo
 \fi
}%
\providecommand \@ifx [1]{%
 \ifx #1\expandafter \@firstoftwo
 \else \expandafter \@secondoftwo
 \fi
}%
\providecommand \natexlab [1]{#1}%
\providecommand \enquote  [1]{``#1''}%
\providecommand \bibnamefont  [1]{#1}%
\providecommand \bibfnamefont [1]{#1}%
\providecommand \citenamefont [1]{#1}%
\providecommand \href@noop [0]{\@secondoftwo}%
\providecommand \href [0]{\begingroup \@sanitize@url \@href}%
\providecommand \@href[1]{\@@startlink{#1}\@@href}%
\providecommand \@@href[1]{\endgroup#1\@@endlink}%
\providecommand \@sanitize@url [0]{\catcode `\\12\catcode `\$12\catcode
  `\&12\catcode `\#12\catcode `\^12\catcode `\_12\catcode `\%12\relax}%
\providecommand \@@startlink[1]{}%
\providecommand \@@endlink[0]{}%
\providecommand \url  [0]{\begingroup\@sanitize@url \@url }%
\providecommand \@url [1]{\endgroup\@href {#1}{\urlprefix }}%
\providecommand \urlprefix  [0]{URL }%
\providecommand \Eprint [0]{\href }%
\providecommand \doibase [0]{http://dx.doi.org/}%
\providecommand \selectlanguage [0]{\@gobble}%
\providecommand \bibinfo  [0]{\@secondoftwo}%
\providecommand \bibfield  [0]{\@secondoftwo}%
\providecommand \translation [1]{[#1]}%
\providecommand \BibitemOpen [0]{}%
\providecommand \bibitemStop [0]{}%
\providecommand \bibitemNoStop [0]{.\EOS\space}%
\providecommand \EOS [0]{\spacefactor3000\relax}%
\providecommand \BibitemShut  [1]{\csname bibitem#1\endcsname}%
\let\auto@bib@innerbib\@empty
\bibitem [{\citenamefont {Wang}\ \emph {et~al.}(2009)\citenamefont {Wang},
  \citenamefont {Chong}, \citenamefont {Joannopoulos},\ and\ \citenamefont
  {Solja\v{c}i\'{c}}}]{Wang2009}%
  \BibitemOpen
  \bibfield  {author} {\bibinfo {author} {\bibfnamefont {Z.}~\bibnamefont
  {Wang}}, \bibinfo {author} {\bibfnamefont {Y.}~\bibnamefont {Chong}},
  \bibinfo {author} {\bibfnamefont {J.~D.}\ \bibnamefont {Joannopoulos}}, \
  and\ \bibinfo {author} {\bibfnamefont {M.}~\bibnamefont {Solja\v{c}i\'{c}}},\
  }\href {http://dx.doi.org/10.1038/nature08293} {\ \textbf {\bibinfo {volume}
  {461}},\ \bibinfo {pages} {772 EP } (\bibinfo {year} {2009})},\ \bibinfo
  {note} {article}\BibitemShut {NoStop}%
\bibitem [{\citenamefont {Poo}\ \emph {et~al.}(2011)\citenamefont {Poo},
  \citenamefont {Wu}, \citenamefont {Lin}, \citenamefont {Yang},\ and\
  \citenamefont {Chan}}]{PhysRevLett.106.093903}%
  \BibitemOpen
  \bibfield  {author} {\bibinfo {author} {\bibfnamefont {Y.}~\bibnamefont
  {Poo}}, \bibinfo {author} {\bibfnamefont {R.-x.}\ \bibnamefont {Wu}},
  \bibinfo {author} {\bibfnamefont {Z.}~\bibnamefont {Lin}}, \bibinfo {author}
  {\bibfnamefont {Y.}~\bibnamefont {Yang}}, \ and\ \bibinfo {author}
  {\bibfnamefont {C.~T.}\ \bibnamefont {Chan}},\ }\href {\doibase
  10.1103/PhysRevLett.106.093903} {\bibfield  {journal} {\bibinfo  {journal}
  {Phys. Rev. Lett.}\ }\textbf {\bibinfo {volume} {106}},\ \bibinfo {pages}
  {093903} (\bibinfo {year} {2011})}\BibitemShut {NoStop}%
\bibitem [{\citenamefont {Hafezi}\ \emph {et~al.}(2013)\citenamefont {Hafezi},
  \citenamefont {Mittal}, \citenamefont {Fan}, \citenamefont {Migdall},\ and\
  \citenamefont {Taylor}}]{Hafezi2013}%
  \BibitemOpen
  \bibfield  {author} {\bibinfo {author} {\bibfnamefont {M.}~\bibnamefont
  {Hafezi}}, \bibinfo {author} {\bibfnamefont {S.}~\bibnamefont {Mittal}},
  \bibinfo {author} {\bibfnamefont {J.}~\bibnamefont {Fan}}, \bibinfo {author}
  {\bibfnamefont {A.}~\bibnamefont {Migdall}}, \ and\ \bibinfo {author}
  {\bibfnamefont {J.~M.}\ \bibnamefont {Taylor}},\ }\href
  {http://dx.doi.org/10.1038/nphoton.2013.274} {\bibfield  {journal} {\bibinfo
  {journal} {Nature Photonics}\ }\textbf {\bibinfo {volume} {7}},\ \bibinfo
  {pages} {1001 EP } (\bibinfo {year} {2013})},\ \bibinfo {note}
  {article}\BibitemShut {NoStop}%
\bibitem [{\citenamefont {Mittal}\ \emph {et~al.}(2014)\citenamefont {Mittal},
  \citenamefont {Fan}, \citenamefont {Faez}, \citenamefont {Migdall},
  \citenamefont {Taylor},\ and\ \citenamefont
  {Hafezi}}]{PhysRevLett.113.087403}%
  \BibitemOpen
  \bibfield  {author} {\bibinfo {author} {\bibfnamefont {S.}~\bibnamefont
  {Mittal}}, \bibinfo {author} {\bibfnamefont {J.}~\bibnamefont {Fan}},
  \bibinfo {author} {\bibfnamefont {S.}~\bibnamefont {Faez}}, \bibinfo {author}
  {\bibfnamefont {A.}~\bibnamefont {Migdall}}, \bibinfo {author} {\bibfnamefont
  {J.~M.}\ \bibnamefont {Taylor}}, \ and\ \bibinfo {author} {\bibfnamefont
  {M.}~\bibnamefont {Hafezi}},\ }\href {\doibase
  10.1103/PhysRevLett.113.087403} {\bibfield  {journal} {\bibinfo  {journal}
  {Phys. Rev. Lett.}\ }\textbf {\bibinfo {volume} {113}},\ \bibinfo {pages}
  {087403} (\bibinfo {year} {2014})}\BibitemShut {NoStop}%
\bibitem [{\citenamefont {Mittal}\ \emph {et~al.}(2016)\citenamefont {Mittal},
  \citenamefont {Ganeshan}, \citenamefont {Fan}, \citenamefont {Vaezi},\ and\
  \citenamefont {Hafezi}}]{Mittal2016}%
  \BibitemOpen
  \bibfield  {author} {\bibinfo {author} {\bibfnamefont {S.}~\bibnamefont
  {Mittal}}, \bibinfo {author} {\bibfnamefont {S.}~\bibnamefont {Ganeshan}},
  \bibinfo {author} {\bibfnamefont {J.}~\bibnamefont {Fan}}, \bibinfo {author}
  {\bibfnamefont {A.}~\bibnamefont {Vaezi}}, \ and\ \bibinfo {author}
  {\bibfnamefont {M.}~\bibnamefont {Hafezi}},\ }\href
  {http://dx.doi.org/10.1038/nphoton.2016.10} {\bibfield  {journal} {\bibinfo
  {journal} {Nature Photonics}\ }\textbf {\bibinfo {volume} {10}},\ \bibinfo
  {pages} {180 EP } (\bibinfo {year} {2016})}\BibitemShut {NoStop}%
\bibitem [{\citenamefont {Chen}\ \emph {et~al.}(2014)\citenamefont {Chen},
  \citenamefont {Jiang}, \citenamefont {Chen}, \citenamefont {Zhu},
  \citenamefont {Zhou}, \citenamefont {Dong},\ and\ \citenamefont
  {Chan}}]{Chen2014}%
  \BibitemOpen
  \bibfield  {author} {\bibinfo {author} {\bibfnamefont {W.-J.}\ \bibnamefont
  {Chen}}, \bibinfo {author} {\bibfnamefont {S.-J.}\ \bibnamefont {Jiang}},
  \bibinfo {author} {\bibfnamefont {X.-D.}\ \bibnamefont {Chen}}, \bibinfo
  {author} {\bibfnamefont {B.}~\bibnamefont {Zhu}}, \bibinfo {author}
  {\bibfnamefont {L.}~\bibnamefont {Zhou}}, \bibinfo {author} {\bibfnamefont
  {J.-W.}\ \bibnamefont {Dong}}, \ and\ \bibinfo {author} {\bibfnamefont
  {C.~T.}\ \bibnamefont {Chan}},\ }\href {http://dx.doi.org/10.1038/ncomms6782}
  {\bibfield  {journal} {\bibinfo  {journal} {Nature Communications}\ }\textbf
  {\bibinfo {volume} {5}},\ \bibinfo {pages} {5782 EP } (\bibinfo {year}
  {2014})},\ \bibinfo {note} {article}\BibitemShut {NoStop}%
\bibitem [{\citenamefont {Slobozhanyuk}\ \emph {et~al.}(2016)\citenamefont
  {Slobozhanyuk}, \citenamefont {Khanikaev}, \citenamefont {Filonov},
  \citenamefont {Smirnova}, \citenamefont {Miroshnichenko},\ and\ \citenamefont
  {Kivshar}}]{Slobozhanyuk2016}%
  \BibitemOpen
  \bibfield  {author} {\bibinfo {author} {\bibfnamefont {A.~P.}\ \bibnamefont
  {Slobozhanyuk}}, \bibinfo {author} {\bibfnamefont {A.~B.}\ \bibnamefont
  {Khanikaev}}, \bibinfo {author} {\bibfnamefont {D.~S.}\ \bibnamefont
  {Filonov}}, \bibinfo {author} {\bibfnamefont {D.~A.}\ \bibnamefont
  {Smirnova}}, \bibinfo {author} {\bibfnamefont {A.~E.}\ \bibnamefont
  {Miroshnichenko}}, \ and\ \bibinfo {author} {\bibfnamefont {Y.~S.}\
  \bibnamefont {Kivshar}},\ }\href {http://dx.doi.org/10.1038/srep22270}
  {\bibfield  {journal} {\bibinfo  {journal} {Scientific Reports}\ }\textbf
  {\bibinfo {volume} {6}},\ \bibinfo {pages} {22270 EP } (\bibinfo {year}
  {2016})},\ \bibinfo {note} {article}\BibitemShut {NoStop}%
\bibitem [{\citenamefont {Poli}\ \emph {et~al.}(2015)\citenamefont {Poli},
  \citenamefont {Bellec}, \citenamefont {Kuhl}, \citenamefont {Mortessagne},\
  and\ \citenamefont {Schomerus}}]{Poli2015}%
  \BibitemOpen
  \bibfield  {author} {\bibinfo {author} {\bibfnamefont {C.}~\bibnamefont
  {Poli}}, \bibinfo {author} {\bibfnamefont {M.}~\bibnamefont {Bellec}},
  \bibinfo {author} {\bibfnamefont {U.}~\bibnamefont {Kuhl}}, \bibinfo {author}
  {\bibfnamefont {F.}~\bibnamefont {Mortessagne}}, \ and\ \bibinfo {author}
  {\bibfnamefont {H.}~\bibnamefont {Schomerus}},\ }\href
  {http://dx.doi.org/10.1038/ncomms7710} {\bibfield  {journal} {\bibinfo
  {journal} {Nature Communications}\ }\textbf {\bibinfo {volume} {6}},\
  \bibinfo {pages} {6710 EP } (\bibinfo {year} {2015})},\ \bibinfo {note}
  {article}\BibitemShut {NoStop}%
\bibitem [{\citenamefont {Zeuner}\ \emph {et~al.}(2015)\citenamefont {Zeuner},
  \citenamefont {Rechtsman}, \citenamefont {Plotnik}, \citenamefont {Lumer},
  \citenamefont {Nolte}, \citenamefont {Rudner}, \citenamefont {Segev},\ and\
  \citenamefont {Szameit}}]{PhysRevLett.115.040402}%
  \BibitemOpen
  \bibfield  {author} {\bibinfo {author} {\bibfnamefont {J.~M.}\ \bibnamefont
  {Zeuner}}, \bibinfo {author} {\bibfnamefont {M.~C.}\ \bibnamefont
  {Rechtsman}}, \bibinfo {author} {\bibfnamefont {Y.}~\bibnamefont {Plotnik}},
  \bibinfo {author} {\bibfnamefont {Y.}~\bibnamefont {Lumer}}, \bibinfo
  {author} {\bibfnamefont {S.}~\bibnamefont {Nolte}}, \bibinfo {author}
  {\bibfnamefont {M.~S.}\ \bibnamefont {Rudner}}, \bibinfo {author}
  {\bibfnamefont {M.}~\bibnamefont {Segev}}, \ and\ \bibinfo {author}
  {\bibfnamefont {A.}~\bibnamefont {Szameit}},\ }\href {\doibase
  10.1103/PhysRevLett.115.040402} {\bibfield  {journal} {\bibinfo  {journal}
  {Phys. Rev. Lett.}\ }\textbf {\bibinfo {volume} {115}},\ \bibinfo {pages}
  {040402} (\bibinfo {year} {2015})}\BibitemShut {NoStop}%
\bibitem [{\citenamefont {Leykam}\ and\ \citenamefont
  {Chong}(2016)}]{PhysRevLett.117.143901}%
  \BibitemOpen
  \bibfield  {author} {\bibinfo {author} {\bibfnamefont {D.}~\bibnamefont
  {Leykam}}\ and\ \bibinfo {author} {\bibfnamefont {Y.~D.}\ \bibnamefont
  {Chong}},\ }\href {\doibase 10.1103/PhysRevLett.117.143901} {\bibfield
  {journal} {\bibinfo  {journal} {Phys. Rev. Lett.}\ }\textbf {\bibinfo
  {volume} {117}},\ \bibinfo {pages} {143901} (\bibinfo {year}
  {2016})}\BibitemShut {NoStop}%
\bibitem [{\citenamefont {Zhou}\ \emph {et~al.}(2017)\citenamefont {Zhou},
  \citenamefont {Wang}, \citenamefont {Leykam},\ and\ \citenamefont
  {Chong}}]{1367-2630-19-9-095002}%
  \BibitemOpen
  \bibfield  {author} {\bibinfo {author} {\bibfnamefont {X.}~\bibnamefont
  {Zhou}}, \bibinfo {author} {\bibfnamefont {Y.}~\bibnamefont {Wang}}, \bibinfo
  {author} {\bibfnamefont {D.}~\bibnamefont {Leykam}}, \ and\ \bibinfo {author}
  {\bibfnamefont {Y.~D.}\ \bibnamefont {Chong}},\ }\href
  {http://stacks.iop.org/1367-2630/19/i=9/a=095002} {\bibfield  {journal}
  {\bibinfo  {journal} {New Journal of Physics}\ }\textbf {\bibinfo {volume}
  {19}},\ \bibinfo {pages} {095002} (\bibinfo {year} {2017})}\BibitemShut
  {NoStop}%
\bibitem [{\citenamefont {Barik}\ \emph {et~al.}(2018)\citenamefont {Barik},
  \citenamefont {Karasahin}, \citenamefont {Flower}, \citenamefont {Cai},
  \citenamefont {Miyake}, \citenamefont {DeGottardi}, \citenamefont {Hafezi},\
  and\ \citenamefont {Waks}}]{Barik666}%
  \BibitemOpen
  \bibfield  {author} {\bibinfo {author} {\bibfnamefont {S.}~\bibnamefont
  {Barik}}, \bibinfo {author} {\bibfnamefont {A.}~\bibnamefont {Karasahin}},
  \bibinfo {author} {\bibfnamefont {C.}~\bibnamefont {Flower}}, \bibinfo
  {author} {\bibfnamefont {T.}~\bibnamefont {Cai}}, \bibinfo {author}
  {\bibfnamefont {H.}~\bibnamefont {Miyake}}, \bibinfo {author} {\bibfnamefont
  {W.}~\bibnamefont {DeGottardi}}, \bibinfo {author} {\bibfnamefont
  {M.}~\bibnamefont {Hafezi}}, \ and\ \bibinfo {author} {\bibfnamefont
  {E.}~\bibnamefont {Waks}},\ }\href {\doibase 10.1126/science.aaq0327}
  {\bibfield  {journal} {\bibinfo  {journal} {Science}\ }\textbf {\bibinfo
  {volume} {359}},\ \bibinfo {pages} {666} (\bibinfo {year} {2018})},\ \Eprint
  {http://arxiv.org/abs/http://science.sciencemag.org/content/359/6376/666.full.pdf}
  {http://science.sciencemag.org/content/359/6376/666.full.pdf} \BibitemShut
  {NoStop}%
\bibitem [{\citenamefont {Khanikaev}\ \emph {et~al.}(2012)\citenamefont
  {Khanikaev}, \citenamefont {Hossein~Mousavi}, \citenamefont {Tse},
  \citenamefont {Kargarian}, \citenamefont {MacDonald},\ and\ \citenamefont
  {Shvets}}]{Khanikaev2012}%
  \BibitemOpen
  \bibfield  {author} {\bibinfo {author} {\bibfnamefont {A.~B.}\ \bibnamefont
  {Khanikaev}}, \bibinfo {author} {\bibfnamefont {S.}~\bibnamefont
  {Hossein~Mousavi}}, \bibinfo {author} {\bibfnamefont {W.-K.}\ \bibnamefont
  {Tse}}, \bibinfo {author} {\bibfnamefont {M.}~\bibnamefont {Kargarian}},
  \bibinfo {author} {\bibfnamefont {A.~H.}\ \bibnamefont {MacDonald}}, \ and\
  \bibinfo {author} {\bibfnamefont {G.}~\bibnamefont {Shvets}},\ }\href
  {http://dx.doi.org/10.1038/nmat3520} {\bibfield  {journal} {\bibinfo
  {journal} {Nature Materials}\ }\textbf {\bibinfo {volume} {12}},\ \bibinfo
  {pages} {233 EP } (\bibinfo {year} {2012})},\ \bibinfo {note}
  {article}\BibitemShut {NoStop}%
\bibitem [{\citenamefont {Solnyshkov}\ and\ \citenamefont
  {Malpuech}(2016)}]{SOLNYSHKOV2016920}%
  \BibitemOpen
  \bibfield  {author} {\bibinfo {author} {\bibfnamefont {D.}~\bibnamefont
  {Solnyshkov}}\ and\ \bibinfo {author} {\bibfnamefont {G.}~\bibnamefont
  {Malpuech}},\ }\href {\doibase https://doi.org/10.1016/j.crhy.2016.07.003}
  {\bibfield  {journal} {\bibinfo  {journal} {Comptes Rendus Physique}\
  }\textbf {\bibinfo {volume} {17}},\ \bibinfo {pages} {920 } (\bibinfo {year}
  {2016})},\ \bibinfo {note} {polariton physics / Physique des
  polaritons}\BibitemShut {NoStop}%
\bibitem [{\citenamefont {Nalitov}\ \emph
  {et~al.}(2015{\natexlab{a}})\citenamefont {Nalitov}, \citenamefont
  {Malpuech}, \citenamefont {Ter\ifmmode~\mbox{\c{c}}\else \c{c}\fi{}as},\ and\
  \citenamefont {Solnyshkov}}]{PhysRevLett.114.026803}%
  \BibitemOpen
  \bibfield  {author} {\bibinfo {author} {\bibfnamefont {A.~V.}\ \bibnamefont
  {Nalitov}}, \bibinfo {author} {\bibfnamefont {G.}~\bibnamefont {Malpuech}},
  \bibinfo {author} {\bibfnamefont {H.}~\bibnamefont
  {Ter\ifmmode~\mbox{\c{c}}\else \c{c}\fi{}as}}, \ and\ \bibinfo {author}
  {\bibfnamefont {D.~D.}\ \bibnamefont {Solnyshkov}},\ }\href {\doibase
  10.1103/PhysRevLett.114.026803} {\bibfield  {journal} {\bibinfo  {journal}
  {Phys. Rev. Lett.}\ }\textbf {\bibinfo {volume} {114}},\ \bibinfo {pages}
  {026803} (\bibinfo {year} {2015}{\natexlab{a}})}\BibitemShut {NoStop}%
\bibitem [{\citenamefont {Aidelsburger}\ \emph {et~al.}(2018)\citenamefont
  {Aidelsburger}, \citenamefont {Nascimbene},\ and\ \citenamefont
  {Goldman}}]{AIDELSBURGER2018}%
  \BibitemOpen
  \bibfield  {author} {\bibinfo {author} {\bibfnamefont {M.}~\bibnamefont
  {Aidelsburger}}, \bibinfo {author} {\bibfnamefont {S.}~\bibnamefont
  {Nascimbene}}, \ and\ \bibinfo {author} {\bibfnamefont {N.}~\bibnamefont
  {Goldman}},\ }\href {\doibase https://doi.org/10.1016/j.crhy.2018.03.002}
  {\bibfield  {journal} {\bibinfo  {journal} {Comptes Rendus Physique}\ }
  (\bibinfo {year} {2018}),\
  https://doi.org/10.1016/j.crhy.2018.03.002}\BibitemShut {NoStop}%
\bibitem [{\citenamefont {Nalitov}\ \emph
  {et~al.}(2015{\natexlab{b}})\citenamefont {Nalitov}, \citenamefont
  {Solnyshkov},\ and\ \citenamefont {Malpuech}}]{PhysRevLett.114.116401}%
  \BibitemOpen
  \bibfield  {author} {\bibinfo {author} {\bibfnamefont {A.~V.}\ \bibnamefont
  {Nalitov}}, \bibinfo {author} {\bibfnamefont {D.~D.}\ \bibnamefont
  {Solnyshkov}}, \ and\ \bibinfo {author} {\bibfnamefont {G.}~\bibnamefont
  {Malpuech}},\ }\href {\doibase 10.1103/PhysRevLett.114.116401} {\bibfield
  {journal} {\bibinfo  {journal} {Phys. Rev. Lett.}\ }\textbf {\bibinfo
  {volume} {114}},\ \bibinfo {pages} {116401} (\bibinfo {year}
  {2015}{\natexlab{b}})}\BibitemShut {NoStop}%
\bibitem [{\citenamefont {Gulevich}\ \emph {et~al.}(2017)\citenamefont
  {Gulevich}, \citenamefont {Yudin}, \citenamefont {Skryabin}, \citenamefont
  {Iorsh},\ and\ \citenamefont {Shelykh}}]{Gulevich2017}%
  \BibitemOpen
  \bibfield  {author} {\bibinfo {author} {\bibfnamefont {D.~R.}\ \bibnamefont
  {Gulevich}}, \bibinfo {author} {\bibfnamefont {D.}~\bibnamefont {Yudin}},
  \bibinfo {author} {\bibfnamefont {D.~V.}\ \bibnamefont {Skryabin}}, \bibinfo
  {author} {\bibfnamefont {I.~V.}\ \bibnamefont {Iorsh}}, \ and\ \bibinfo
  {author} {\bibfnamefont {I.~A.}\ \bibnamefont {Shelykh}},\ }\href {\doibase
  10.1038/s41598-017-01646-y} {\bibfield  {journal} {\bibinfo  {journal}
  {Scientific Reports}\ }\textbf {\bibinfo {volume} {7}},\ \bibinfo {pages}
  {1780} (\bibinfo {year} {2017})}\BibitemShut {NoStop}%
\bibitem [{\citenamefont {Li}\ \emph {et~al.}(2018)\citenamefont {Li},
  \citenamefont {Ye}, \citenamefont {Chen}, \citenamefont {Kartashov},
  \citenamefont {Ferrando}, \citenamefont {Torner},\ and\ \citenamefont
  {Skryabin}}]{PhysRevB.97.081103}%
  \BibitemOpen
  \bibfield  {author} {\bibinfo {author} {\bibfnamefont {C.}~\bibnamefont
  {Li}}, \bibinfo {author} {\bibfnamefont {F.}~\bibnamefont {Ye}}, \bibinfo
  {author} {\bibfnamefont {X.}~\bibnamefont {Chen}}, \bibinfo {author}
  {\bibfnamefont {Y.~V.}\ \bibnamefont {Kartashov}}, \bibinfo {author}
  {\bibfnamefont {A.}~\bibnamefont {Ferrando}}, \bibinfo {author}
  {\bibfnamefont {L.}~\bibnamefont {Torner}}, \ and\ \bibinfo {author}
  {\bibfnamefont {D.~V.}\ \bibnamefont {Skryabin}},\ }\href {\doibase
  10.1103/PhysRevB.97.081103} {\bibfield  {journal} {\bibinfo  {journal} {Phys.
  Rev. B}\ }\textbf {\bibinfo {volume} {97}},\ \bibinfo {pages} {081103}
  (\bibinfo {year} {2018})}\BibitemShut {NoStop}%
\bibitem [{\citenamefont {Solnyshkov}\ \emph
  {et~al.}(2016{\natexlab{a}})\citenamefont {Solnyshkov}, \citenamefont
  {Nalitov}, \citenamefont {Teklu}, \citenamefont {Franck},\ and\ \citenamefont
  {Malpuech}}]{PhysRevB.93.085404}%
  \BibitemOpen
  \bibfield  {author} {\bibinfo {author} {\bibfnamefont {D.}~\bibnamefont
  {Solnyshkov}}, \bibinfo {author} {\bibfnamefont {A.}~\bibnamefont {Nalitov}},
  \bibinfo {author} {\bibfnamefont {B.}~\bibnamefont {Teklu}}, \bibinfo
  {author} {\bibfnamefont {L.}~\bibnamefont {Franck}}, \ and\ \bibinfo {author}
  {\bibfnamefont {G.}~\bibnamefont {Malpuech}},\ }\href {\doibase
  10.1103/PhysRevB.93.085404} {\bibfield  {journal} {\bibinfo  {journal} {Phys.
  Rev. B}\ }\textbf {\bibinfo {volume} {93}},\ \bibinfo {pages} {085404}
  (\bibinfo {year} {2016}{\natexlab{a}})}\BibitemShut {NoStop}%
\bibitem [{\citenamefont {Sala}\ \emph {et~al.}(2015)\citenamefont {Sala},
  \citenamefont {Solnyshkov}, \citenamefont {Carusotto}, \citenamefont
  {Jacqmin}, \citenamefont {Lema\^{\i}tre}, \citenamefont
  {Ter\ifmmode~\mbox{\c{c}}\else \c{c}\fi{}as}, \citenamefont {Nalitov},
  \citenamefont {Abbarchi}, \citenamefont {Galopin}, \citenamefont {Sagnes},
  \citenamefont {Bloch}, \citenamefont {Malpuech},\ and\ \citenamefont
  {Amo}}]{PhysRevX.5.011034}%
  \BibitemOpen
  \bibfield  {author} {\bibinfo {author} {\bibfnamefont {V.~G.}\ \bibnamefont
  {Sala}}, \bibinfo {author} {\bibfnamefont {D.~D.}\ \bibnamefont
  {Solnyshkov}}, \bibinfo {author} {\bibfnamefont {I.}~\bibnamefont
  {Carusotto}}, \bibinfo {author} {\bibfnamefont {T.}~\bibnamefont {Jacqmin}},
  \bibinfo {author} {\bibfnamefont {A.}~\bibnamefont {Lema\^{\i}tre}}, \bibinfo
  {author} {\bibfnamefont {H.}~\bibnamefont {Ter\ifmmode~\mbox{\c{c}}\else
  \c{c}\fi{}as}}, \bibinfo {author} {\bibfnamefont {A.}~\bibnamefont
  {Nalitov}}, \bibinfo {author} {\bibfnamefont {M.}~\bibnamefont {Abbarchi}},
  \bibinfo {author} {\bibfnamefont {E.}~\bibnamefont {Galopin}}, \bibinfo
  {author} {\bibfnamefont {I.}~\bibnamefont {Sagnes}}, \bibinfo {author}
  {\bibfnamefont {J.}~\bibnamefont {Bloch}}, \bibinfo {author} {\bibfnamefont
  {G.}~\bibnamefont {Malpuech}}, \ and\ \bibinfo {author} {\bibfnamefont
  {A.}~\bibnamefont {Amo}},\ }\href {\doibase 10.1103/PhysRevX.5.011034}
  {\bibfield  {journal} {\bibinfo  {journal} {Phys. Rev. X}\ }\textbf {\bibinfo
  {volume} {5}},\ \bibinfo {pages} {011034} (\bibinfo {year}
  {2015})}\BibitemShut {NoStop}%
\bibitem [{\citenamefont {Whittaker}\ \emph {et~al.}(2018)\citenamefont
  {Whittaker}, \citenamefont {Cancellieri}, \citenamefont {Walker},
  \citenamefont {Gulevich}, \citenamefont {Schomerus}, \citenamefont
  {Vaitiekus}, \citenamefont {Royall}, \citenamefont {Whittaker}, \citenamefont
  {Clarke}, \citenamefont {Iorsh}, \citenamefont {Shelykh}, \citenamefont
  {Skolnick},\ and\ \citenamefont {Krizhanovskii}}]{PhysRevLett.120.097401}%
  \BibitemOpen
  \bibfield  {author} {\bibinfo {author} {\bibfnamefont {C.~E.}\ \bibnamefont
  {Whittaker}}, \bibinfo {author} {\bibfnamefont {E.}~\bibnamefont
  {Cancellieri}}, \bibinfo {author} {\bibfnamefont {P.~M.}\ \bibnamefont
  {Walker}}, \bibinfo {author} {\bibfnamefont {D.~R.}\ \bibnamefont
  {Gulevich}}, \bibinfo {author} {\bibfnamefont {H.}~\bibnamefont {Schomerus}},
  \bibinfo {author} {\bibfnamefont {D.}~\bibnamefont {Vaitiekus}}, \bibinfo
  {author} {\bibfnamefont {B.}~\bibnamefont {Royall}}, \bibinfo {author}
  {\bibfnamefont {D.~M.}\ \bibnamefont {Whittaker}}, \bibinfo {author}
  {\bibfnamefont {E.}~\bibnamefont {Clarke}}, \bibinfo {author} {\bibfnamefont
  {I.~V.}\ \bibnamefont {Iorsh}}, \bibinfo {author} {\bibfnamefont {I.~A.}\
  \bibnamefont {Shelykh}}, \bibinfo {author} {\bibfnamefont {M.~S.}\
  \bibnamefont {Skolnick}}, \ and\ \bibinfo {author} {\bibfnamefont {D.~N.}\
  \bibnamefont {Krizhanovskii}},\ }\href {\doibase
  10.1103/PhysRevLett.120.097401} {\bibfield  {journal} {\bibinfo  {journal}
  {Phys. Rev. Lett.}\ }\textbf {\bibinfo {volume} {120}},\ \bibinfo {pages}
  {097401} (\bibinfo {year} {2018})}\BibitemShut {NoStop}%
\bibitem [{\citenamefont {Klembt}\ \emph {et~al.}(2017)\citenamefont {Klembt},
  \citenamefont {Harder}, \citenamefont {Egorov}, \citenamefont {Winkler},
  \citenamefont {Suchomel}, \citenamefont {Beierlein}, \citenamefont
  {Emmerling}, \citenamefont {Schneider},\ and\ \citenamefont
  {Höfling}}]{doi:10.1063/1.4995385}%
  \BibitemOpen
  \bibfield  {author} {\bibinfo {author} {\bibfnamefont {S.}~\bibnamefont
  {Klembt}}, \bibinfo {author} {\bibfnamefont {T.~H.}\ \bibnamefont {Harder}},
  \bibinfo {author} {\bibfnamefont {O.~A.}\ \bibnamefont {Egorov}}, \bibinfo
  {author} {\bibfnamefont {K.}~\bibnamefont {Winkler}}, \bibinfo {author}
  {\bibfnamefont {H.}~\bibnamefont {Suchomel}}, \bibinfo {author}
  {\bibfnamefont {J.}~\bibnamefont {Beierlein}}, \bibinfo {author}
  {\bibfnamefont {M.}~\bibnamefont {Emmerling}}, \bibinfo {author}
  {\bibfnamefont {C.}~\bibnamefont {Schneider}}, \ and\ \bibinfo {author}
  {\bibfnamefont {S.}~\bibnamefont {Höfling}},\ }\href {\doibase
  10.1063/1.4995385} {\bibfield  {journal} {\bibinfo  {journal} {Applied
  Physics Letters}\ }\textbf {\bibinfo {volume} {111}},\ \bibinfo {pages}
  {231102} (\bibinfo {year} {2017})},\ \Eprint
  {http://arxiv.org/abs/https://doi.org/10.1063/1.4995385}
  {https://doi.org/10.1063/1.4995385} \BibitemShut {NoStop}%
\bibitem [{\citenamefont {Su}\ \emph {et~al.}(1979)\citenamefont {Su},
  \citenamefont {Schrieffer},\ and\ \citenamefont
  {Heeger}}]{PhysRevLett.42.1698}%
  \BibitemOpen
  \bibfield  {author} {\bibinfo {author} {\bibfnamefont {W.~P.}\ \bibnamefont
  {Su}}, \bibinfo {author} {\bibfnamefont {J.~R.}\ \bibnamefont {Schrieffer}},
  \ and\ \bibinfo {author} {\bibfnamefont {A.~J.}\ \bibnamefont {Heeger}},\
  }\href {\doibase 10.1103/PhysRevLett.42.1698} {\bibfield  {journal} {\bibinfo
   {journal} {Phys. Rev. Lett.}\ }\textbf {\bibinfo {volume} {42}},\ \bibinfo
  {pages} {1698} (\bibinfo {year} {1979})}\BibitemShut {NoStop}%
\bibitem [{\citenamefont {Malkova}\ \emph {et~al.}(2009)\citenamefont
  {Malkova}, \citenamefont {Hromada}, \citenamefont {Wang}, \citenamefont
  {Bryant},\ and\ \citenamefont {Chen}}]{Malkova:09}%
  \BibitemOpen
  \bibfield  {author} {\bibinfo {author} {\bibfnamefont {N.}~\bibnamefont
  {Malkova}}, \bibinfo {author} {\bibfnamefont {I.}~\bibnamefont {Hromada}},
  \bibinfo {author} {\bibfnamefont {X.}~\bibnamefont {Wang}}, \bibinfo {author}
  {\bibfnamefont {G.}~\bibnamefont {Bryant}}, \ and\ \bibinfo {author}
  {\bibfnamefont {Z.}~\bibnamefont {Chen}},\ }\href {\doibase
  10.1364/OL.34.001633} {\bibfield  {journal} {\bibinfo  {journal} {Opt.
  Lett.}\ }\textbf {\bibinfo {volume} {34}},\ \bibinfo {pages} {1633} (\bibinfo
  {year} {2009})}\BibitemShut {NoStop}%
\bibitem [{\citenamefont {Cheng}\ \emph {et~al.}(2015)\citenamefont {Cheng},
  \citenamefont {Pan}, \citenamefont {Wang}, \citenamefont {Li},\ and\
  \citenamefont {Zhu}}]{LPOR:LPOR201400462}%
  \BibitemOpen
  \bibfield  {author} {\bibinfo {author} {\bibfnamefont {Q.}~\bibnamefont
  {Cheng}}, \bibinfo {author} {\bibfnamefont {Y.}~\bibnamefont {Pan}}, \bibinfo
  {author} {\bibfnamefont {Q.}~\bibnamefont {Wang}}, \bibinfo {author}
  {\bibfnamefont {T.}~\bibnamefont {Li}}, \ and\ \bibinfo {author}
  {\bibfnamefont {S.}~\bibnamefont {Zhu}},\ }\href {\doibase
  10.1002/lpor.201400462} {\bibfield  {journal} {\bibinfo  {journal} {Laser \&
  Photonics Reviews}\ }\textbf {\bibinfo {volume} {9}},\ \bibinfo {pages} {392}
  (\bibinfo {year} {2015})}\BibitemShut {NoStop}%
\bibitem [{\citenamefont {Sinev}\ \emph {et~al.}(2015)\citenamefont {Sinev},
  \citenamefont {Mukhin}, \citenamefont {Slobozhanyuk}, \citenamefont
  {Poddubny}, \citenamefont {Miroshnichenko}, \citenamefont {Samusev},\ and\
  \citenamefont {Kivshar}}]{C5NR00231A}%
  \BibitemOpen
  \bibfield  {author} {\bibinfo {author} {\bibfnamefont {I.~S.}\ \bibnamefont
  {Sinev}}, \bibinfo {author} {\bibfnamefont {I.~S.}\ \bibnamefont {Mukhin}},
  \bibinfo {author} {\bibfnamefont {A.~P.}\ \bibnamefont {Slobozhanyuk}},
  \bibinfo {author} {\bibfnamefont {A.~N.}\ \bibnamefont {Poddubny}}, \bibinfo
  {author} {\bibfnamefont {A.~E.}\ \bibnamefont {Miroshnichenko}}, \bibinfo
  {author} {\bibfnamefont {A.~K.}\ \bibnamefont {Samusev}}, \ and\ \bibinfo
  {author} {\bibfnamefont {Y.~S.}\ \bibnamefont {Kivshar}},\ }\href {\doibase
  10.1039/C5NR00231A} {\bibfield  {journal} {\bibinfo  {journal} {Nanoscale}\
  }\textbf {\bibinfo {volume} {7}},\ \bibinfo {pages} {11904} (\bibinfo {year}
  {2015})}\BibitemShut {NoStop}%
\bibitem [{\citenamefont {Weimann}\ \emph {et~al.}(2016)\citenamefont
  {Weimann}, \citenamefont {Kremer}, \citenamefont {Plotnik}, \citenamefont
  {Lumer}, \citenamefont {Nolte}, \citenamefont {Makris}, \citenamefont
  {Segev}, \citenamefont {Rechtsman},\ and\ \citenamefont
  {Szameit}}]{Weimann2016}%
  \BibitemOpen
  \bibfield  {author} {\bibinfo {author} {\bibfnamefont {S.}~\bibnamefont
  {Weimann}}, \bibinfo {author} {\bibfnamefont {M.}~\bibnamefont {Kremer}},
  \bibinfo {author} {\bibfnamefont {Y.}~\bibnamefont {Plotnik}}, \bibinfo
  {author} {\bibfnamefont {Y.}~\bibnamefont {Lumer}}, \bibinfo {author}
  {\bibfnamefont {S.}~\bibnamefont {Nolte}}, \bibinfo {author} {\bibfnamefont
  {K.~G.}\ \bibnamefont {Makris}}, \bibinfo {author} {\bibfnamefont
  {M.}~\bibnamefont {Segev}}, \bibinfo {author} {\bibfnamefont {M.~.~C.}\
  \bibnamefont {Rechtsman}}, \ and\ \bibinfo {author} {\bibfnamefont
  {A.}~\bibnamefont {Szameit}},\ }\href {http://dx.doi.org/10.1038/nmat4811} {\
  \textbf {\bibinfo {volume} {16}},\ \bibinfo {pages} {433 EP } (\bibinfo
  {year} {2016})},\ \bibinfo {note} {article}\BibitemShut {NoStop}%
\bibitem [{\citenamefont {Zhao}\ \emph {et~al.}(2018)\citenamefont {Zhao},
  \citenamefont {Miao}, \citenamefont {Teimourpour}, \citenamefont {Malzard},
  \citenamefont {El-Ganainy}, \citenamefont {Schomerus},\ and\ \citenamefont
  {Feng}}]{Zhao2018}%
  \BibitemOpen
  \bibfield  {author} {\bibinfo {author} {\bibfnamefont {H.}~\bibnamefont
  {Zhao}}, \bibinfo {author} {\bibfnamefont {P.}~\bibnamefont {Miao}}, \bibinfo
  {author} {\bibfnamefont {M.~H.}\ \bibnamefont {Teimourpour}}, \bibinfo
  {author} {\bibfnamefont {S.}~\bibnamefont {Malzard}}, \bibinfo {author}
  {\bibfnamefont {R.}~\bibnamefont {El-Ganainy}}, \bibinfo {author}
  {\bibfnamefont {H.}~\bibnamefont {Schomerus}}, \ and\ \bibinfo {author}
  {\bibfnamefont {L.}~\bibnamefont {Feng}},\ }\href {\doibase
  10.1038/s41467-018-03434-2} {\bibfield  {journal} {\bibinfo  {journal}
  {Nature Communications}\ }\textbf {\bibinfo {volume} {9}},\ \bibinfo {pages}
  {981} (\bibinfo {year} {2018})}\BibitemShut {NoStop}%
\bibitem [{\citenamefont {Parto}\ \emph {et~al.}(2018)\citenamefont {Parto},
  \citenamefont {Wittek}, \citenamefont {Hodaei}, \citenamefont {Harari},
  \citenamefont {Bandres}, \citenamefont {Ren}, \citenamefont {Rechtsman},
  \citenamefont {Segev}, \citenamefont {Christodoulides},\ and\ \citenamefont
  {Khajavikhan}}]{PhysRevLett.120.113901}%
  \BibitemOpen
  \bibfield  {author} {\bibinfo {author} {\bibfnamefont {M.}~\bibnamefont
  {Parto}}, \bibinfo {author} {\bibfnamefont {S.}~\bibnamefont {Wittek}},
  \bibinfo {author} {\bibfnamefont {H.}~\bibnamefont {Hodaei}}, \bibinfo
  {author} {\bibfnamefont {G.}~\bibnamefont {Harari}}, \bibinfo {author}
  {\bibfnamefont {M.~A.}\ \bibnamefont {Bandres}}, \bibinfo {author}
  {\bibfnamefont {J.}~\bibnamefont {Ren}}, \bibinfo {author} {\bibfnamefont
  {M.~C.}\ \bibnamefont {Rechtsman}}, \bibinfo {author} {\bibfnamefont
  {M.}~\bibnamefont {Segev}}, \bibinfo {author} {\bibfnamefont {D.~N.}\
  \bibnamefont {Christodoulides}}, \ and\ \bibinfo {author} {\bibfnamefont
  {M.}~\bibnamefont {Khajavikhan}},\ }\href {\doibase
  10.1103/PhysRevLett.120.113901} {\bibfield  {journal} {\bibinfo  {journal}
  {Phys. Rev. Lett.}\ }\textbf {\bibinfo {volume} {120}},\ \bibinfo {pages}
  {113901} (\bibinfo {year} {2018})}\BibitemShut {NoStop}%
\bibitem [{\citenamefont {Solnyshkov}\ \emph
  {et~al.}(2016{\natexlab{b}})\citenamefont {Solnyshkov}, \citenamefont
  {Nalitov},\ and\ \citenamefont {Malpuech}}]{PhysRevLett.116.046402}%
  \BibitemOpen
  \bibfield  {author} {\bibinfo {author} {\bibfnamefont {D.~D.}\ \bibnamefont
  {Solnyshkov}}, \bibinfo {author} {\bibfnamefont {A.~V.}\ \bibnamefont
  {Nalitov}}, \ and\ \bibinfo {author} {\bibfnamefont {G.}~\bibnamefont
  {Malpuech}},\ }\href {\doibase 10.1103/PhysRevLett.116.046402} {\bibfield
  {journal} {\bibinfo  {journal} {Phys. Rev. Lett.}\ }\textbf {\bibinfo
  {volume} {116}},\ \bibinfo {pages} {046402} (\bibinfo {year}
  {2016}{\natexlab{b}})}\BibitemShut {NoStop}%
\bibitem [{\citenamefont {St-Jean}\ \emph {et~al.}(2017)\citenamefont
  {St-Jean}, \citenamefont {Goblot}, \citenamefont {Galopin}, \citenamefont
  {Lema\^{i}tre}, \citenamefont {Ozawa}, \citenamefont {Le~Gratiet},
  \citenamefont {Sagnes}, \citenamefont {Bloch},\ and\ \citenamefont
  {Amo}}]{St-Jean2017}%
  \BibitemOpen
  \bibfield  {author} {\bibinfo {author} {\bibfnamefont {P.}~\bibnamefont
  {St-Jean}}, \bibinfo {author} {\bibfnamefont {V.}~\bibnamefont {Goblot}},
  \bibinfo {author} {\bibfnamefont {E.}~\bibnamefont {Galopin}}, \bibinfo
  {author} {\bibfnamefont {A.}~\bibnamefont {Lema\^{i}tre}}, \bibinfo {author}
  {\bibfnamefont {T.}~\bibnamefont {Ozawa}}, \bibinfo {author} {\bibfnamefont
  {L.}~\bibnamefont {Le~Gratiet}}, \bibinfo {author} {\bibfnamefont
  {I.}~\bibnamefont {Sagnes}}, \bibinfo {author} {\bibfnamefont
  {J.}~\bibnamefont {Bloch}}, \ and\ \bibinfo {author} {\bibfnamefont
  {A.}~\bibnamefont {Amo}},\ }\href {\doibase 10.1038/s41566-017-0006-2}
  {\bibfield  {journal} {\bibinfo  {journal} {Nature Photonics}\ }\textbf
  {\bibinfo {volume} {11}},\ \bibinfo {pages} {651} (\bibinfo {year}
  {2017})}\BibitemShut {NoStop}%
\bibitem [{\citenamefont {Dufferwiel}\ \emph {et~al.}(2015)\citenamefont
  {Dufferwiel}, \citenamefont {Li}, \citenamefont {Cancellieri}, \citenamefont
  {Giriunas}, \citenamefont {Trichet}, \citenamefont {Whittaker}, \citenamefont
  {Walker}, \citenamefont {Fras}, \citenamefont {Clarke}, \citenamefont
  {Smith}, \citenamefont {Skolnick},\ and\ \citenamefont
  {Krizhanovskii}}]{PhysRevLett.115.246401}%
  \BibitemOpen
  \bibfield  {author} {\bibinfo {author} {\bibfnamefont {S.}~\bibnamefont
  {Dufferwiel}}, \bibinfo {author} {\bibfnamefont {F.}~\bibnamefont {Li}},
  \bibinfo {author} {\bibfnamefont {E.}~\bibnamefont {Cancellieri}}, \bibinfo
  {author} {\bibfnamefont {L.}~\bibnamefont {Giriunas}}, \bibinfo {author}
  {\bibfnamefont {A.~A.~P.}\ \bibnamefont {Trichet}}, \bibinfo {author}
  {\bibfnamefont {D.~M.}\ \bibnamefont {Whittaker}}, \bibinfo {author}
  {\bibfnamefont {P.~M.}\ \bibnamefont {Walker}}, \bibinfo {author}
  {\bibfnamefont {F.}~\bibnamefont {Fras}}, \bibinfo {author} {\bibfnamefont
  {E.}~\bibnamefont {Clarke}}, \bibinfo {author} {\bibfnamefont {J.~M.}\
  \bibnamefont {Smith}}, \bibinfo {author} {\bibfnamefont {M.~S.}\ \bibnamefont
  {Skolnick}}, \ and\ \bibinfo {author} {\bibfnamefont {D.~N.}\ \bibnamefont
  {Krizhanovskii}},\ }\href {\doibase 10.1103/PhysRevLett.115.246401}
  {\bibfield  {journal} {\bibinfo  {journal} {Phys. Rev. Lett.}\ }\textbf
  {\bibinfo {volume} {115}},\ \bibinfo {pages} {246401} (\bibinfo {year}
  {2015})}\BibitemShut {NoStop}%
\bibitem [{\citenamefont {Asb{\'o}th}\ \emph {et~al.}(2016)\citenamefont
  {Asb{\'o}th}, \citenamefont {Oroszl{\'a}ny},\ and\ \citenamefont
  {P{\'a}lyi}}]{asboth2016short}%
  \BibitemOpen
  \bibfield  {author} {\bibinfo {author} {\bibfnamefont {J.}~\bibnamefont
  {Asb{\'o}th}}, \bibinfo {author} {\bibfnamefont {L.}~\bibnamefont
  {Oroszl{\'a}ny}}, \ and\ \bibinfo {author} {\bibfnamefont {A.}~\bibnamefont
  {P{\'a}lyi}},\ }\href {https://books.google.co.uk/books?id=RWKhCwAAQBAJ}
  {\emph {\bibinfo {title} {A Short Course on Topological Insulators: Band
  Structure and Edge States in One and Two Dimensions}}},\ Lecture Notes in
  Physics\ (\bibinfo  {publisher} {Springer International Publishing},\
  \bibinfo {year} {2016})\BibitemShut {NoStop}%
\bibitem [{\citenamefont {Panzarini}\ \emph {et~al.}(1999)\citenamefont
  {Panzarini}, \citenamefont {Andreani}, \citenamefont {Armitage},
  \citenamefont {Baxter}, \citenamefont {Skolnick}, \citenamefont {Astratov},
  \citenamefont {Roberts}, \citenamefont {Kavokin}, \citenamefont
  {Vladimirova},\ and\ \citenamefont {Kaliteevski}}]{PhysRevB.59.5082}%
  \BibitemOpen
  \bibfield  {author} {\bibinfo {author} {\bibfnamefont {G.}~\bibnamefont
  {Panzarini}}, \bibinfo {author} {\bibfnamefont {L.~C.}\ \bibnamefont
  {Andreani}}, \bibinfo {author} {\bibfnamefont {A.}~\bibnamefont {Armitage}},
  \bibinfo {author} {\bibfnamefont {D.}~\bibnamefont {Baxter}}, \bibinfo
  {author} {\bibfnamefont {M.~S.}\ \bibnamefont {Skolnick}}, \bibinfo {author}
  {\bibfnamefont {V.~N.}\ \bibnamefont {Astratov}}, \bibinfo {author}
  {\bibfnamefont {J.~S.}\ \bibnamefont {Roberts}}, \bibinfo {author}
  {\bibfnamefont {A.~V.}\ \bibnamefont {Kavokin}}, \bibinfo {author}
  {\bibfnamefont {M.~R.}\ \bibnamefont {Vladimirova}}, \ and\ \bibinfo {author}
  {\bibfnamefont {M.~A.}\ \bibnamefont {Kaliteevski}},\ }\href {\doibase
  10.1103/PhysRevB.59.5082} {\bibfield  {journal} {\bibinfo  {journal} {Phys.
  Rev. B}\ }\textbf {\bibinfo {volume} {59}},\ \bibinfo {pages} {5082}
  (\bibinfo {year} {1999})}\BibitemShut {NoStop}%
\bibitem [{Note1()}]{Note1}%
  \BibitemOpen
  \bibinfo {note} {See supplementary material at [url] for further details
  about the experimental measurements, additional experimental data from
  several microstructures on the same sample, a description of the
  tight-binding model developed and a discussion of the topological properties
  of the system. Also includes Refs. \cite
  {PhysRevLett.81.2582,1367-2630-12-6-065010}}\BibitemShut {NoStop}%
\bibitem [{\citenamefont {Ryu}\ \emph {et~al.}(2010{\natexlab{a}})\citenamefont
  {Ryu}, \citenamefont {Schnyder}, \citenamefont {Furusaki},\ and\
  \citenamefont {Ludwig}}]{Ryu2010}%
  \BibitemOpen
  \bibfield  {author} {\bibinfo {author} {\bibfnamefont {S.}~\bibnamefont
  {Ryu}}, \bibinfo {author} {\bibfnamefont {A.}~\bibnamefont {Schnyder}},
  \bibinfo {author} {\bibfnamefont {A.}~\bibnamefont {Furusaki}}, \ and\
  \bibinfo {author} {\bibfnamefont {A.}~\bibnamefont {Ludwig}},\ }\href
  {\doibase 10.1088/1367-2630/12/6/065010} {\bibfield  {journal} {\bibinfo
  {journal} {New Journal of Physics}\ }\textbf {\bibinfo {volume} {12}}
  (\bibinfo {year} {2010}{\natexlab{a}}),\ 10.1088/1367-2630/12/6/065010},\
  \bibinfo {note} {cited By 719}\BibitemShut {NoStop}%
\bibitem [{\citenamefont {Shiozaki}\ and\ \citenamefont
  {Sato}(2014)}]{PhysRevB.90.165114}%
  \BibitemOpen
  \bibfield  {author} {\bibinfo {author} {\bibfnamefont {K.}~\bibnamefont
  {Shiozaki}}\ and\ \bibinfo {author} {\bibfnamefont {M.}~\bibnamefont
  {Sato}},\ }\href {\doibase 10.1103/PhysRevB.90.165114} {\bibfield  {journal}
  {\bibinfo  {journal} {Phys. Rev. B}\ }\textbf {\bibinfo {volume} {90}},\
  \bibinfo {pages} {165114} (\bibinfo {year} {2014})}\BibitemShut {NoStop}%
\bibitem [{\citenamefont {Li}\ \emph {et~al.}(2016)\citenamefont {Li},
  \citenamefont {Cancellieri}, \citenamefont {Buonaiuto}, \citenamefont
  {Skolnick}, \citenamefont {Krizhanovskii},\ and\ \citenamefont
  {Whittaker}}]{PhysRevB.94.201301}%
  \BibitemOpen
  \bibfield  {author} {\bibinfo {author} {\bibfnamefont {F.}~\bibnamefont
  {Li}}, \bibinfo {author} {\bibfnamefont {E.}~\bibnamefont {Cancellieri}},
  \bibinfo {author} {\bibfnamefont {G.}~\bibnamefont {Buonaiuto}}, \bibinfo
  {author} {\bibfnamefont {M.~S.}\ \bibnamefont {Skolnick}}, \bibinfo {author}
  {\bibfnamefont {D.~N.}\ \bibnamefont {Krizhanovskii}}, \ and\ \bibinfo
  {author} {\bibfnamefont {D.~M.}\ \bibnamefont {Whittaker}},\ }\href {\doibase
  10.1103/PhysRevB.94.201301} {\bibfield  {journal} {\bibinfo  {journal} {Phys.
  Rev. B}\ }\textbf {\bibinfo {volume} {94}},\ \bibinfo {pages} {201301}
  (\bibinfo {year} {2016})}\BibitemShut {NoStop}%
\bibitem [{\citenamefont {Bayer}\ \emph {et~al.}(1998)\citenamefont {Bayer},
  \citenamefont {Gutbrod}, \citenamefont {Reithmaier}, \citenamefont {Forchel},
  \citenamefont {Reinecke}, \citenamefont {Knipp}, \citenamefont {Dremin},\
  and\ \citenamefont {Kulakovskii}}]{PhysRevLett.81.2582}%
  \BibitemOpen
  \bibfield  {author} {\bibinfo {author} {\bibfnamefont {M.}~\bibnamefont
  {Bayer}}, \bibinfo {author} {\bibfnamefont {T.}~\bibnamefont {Gutbrod}},
  \bibinfo {author} {\bibfnamefont {J.~P.}\ \bibnamefont {Reithmaier}},
  \bibinfo {author} {\bibfnamefont {A.}~\bibnamefont {Forchel}}, \bibinfo
  {author} {\bibfnamefont {T.~L.}\ \bibnamefont {Reinecke}}, \bibinfo {author}
  {\bibfnamefont {P.~A.}\ \bibnamefont {Knipp}}, \bibinfo {author}
  {\bibfnamefont {A.~A.}\ \bibnamefont {Dremin}}, \ and\ \bibinfo {author}
  {\bibfnamefont {V.~D.}\ \bibnamefont {Kulakovskii}},\ }\href {\doibase
  10.1103/PhysRevLett.81.2582} {\bibfield  {journal} {\bibinfo  {journal}
  {Phys. Rev. Lett.}\ }\textbf {\bibinfo {volume} {81}},\ \bibinfo {pages}
  {2582} (\bibinfo {year} {1998})}\BibitemShut {NoStop}%
\bibitem [{\citenamefont {Ryu}\ \emph {et~al.}(2010{\natexlab{b}})\citenamefont
  {Ryu}, \citenamefont {Schnyder}, \citenamefont {Furusaki},\ and\
  \citenamefont {Ludwig}}]{1367-2630-12-6-065010}%
  \BibitemOpen
  \bibfield  {author} {\bibinfo {author} {\bibfnamefont {S.}~\bibnamefont
  {Ryu}}, \bibinfo {author} {\bibfnamefont {A.~P.}\ \bibnamefont {Schnyder}},
  \bibinfo {author} {\bibfnamefont {A.}~\bibnamefont {Furusaki}}, \ and\
  \bibinfo {author} {\bibfnamefont {A.~W.~W.}\ \bibnamefont {Ludwig}},\ }\href
  {http://stacks.iop.org/1367-2630/12/i=6/a=065010} {\bibfield  {journal}
  {\bibinfo  {journal} {New Journal of Physics}\ }\textbf {\bibinfo {volume}
  {12}},\ \bibinfo {pages} {065010} (\bibinfo {year}
  {2010}{\natexlab{b}})}\BibitemShut {NoStop}%
\end{thebibliography}%

\pagebreak
\widetext
\begin{center}
\textbf{\large Supplementary material}
\end{center}
\setcounter{equation}{0}
\setcounter{figure}{0}
\setcounter{table}{0}
\setcounter{page}{1}
\makeatletter
\renewcommand{\theequation}{S\arabic{equation}}
\renewcommand{\thefigure}{S\arabic{figure}}

\section{Experimental setup}

\begin{figure}[h]
\centering
\includegraphics[scale=0.5]{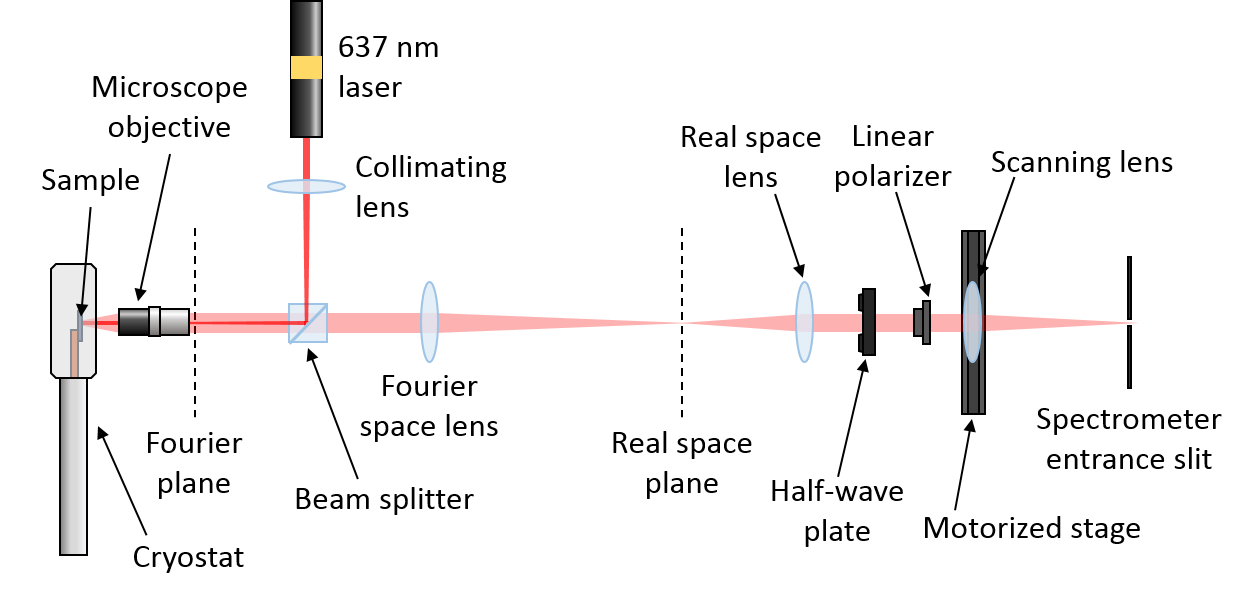}
\caption{Schematic of the reflection $\mu$PL setup used in experiment.}
\label{schema}
\end{figure}

In order to study our photonic structures we hold our sample in a continuous flow cryostat at 8K and perform microphotoluminescence ($\mu$PL) spectroscopy measurements. A schematic of the optical setup is shown in Fig. \ref{schema}. Structures are excited nonresonantly using a cw laser at 637 nm, with an excitation spot which covers a $\sim$20 $\mu$m region on the sample surface. This incoherently populates all the modes of the microstructures. The reflected PL emission is collected with the same 0.60 numerical aperture microscope objective, magnified 30 times and focused onto the entrance slit of a spectrometer connected to a CCD camera. A half-wave plate and linear polarizer are used in the collection path to measure the PL emission in orthogonal polarizations. By scanning the final imaging lens across the spectrometer slit in increments a four-dimensional data set can be constructed, featuring energy, PL intensity, and two spatial dimensions. To filter out parasitic background emission in experimental data, a spatial filter was used in the real space plane to select only PL from the structure being studied, and a long-pass filter was used to suppress scattered laser light.

\section{$P$ orbital SSH chain without polarization}

\begin{figure}[h]
\centering
\includegraphics[scale=1]{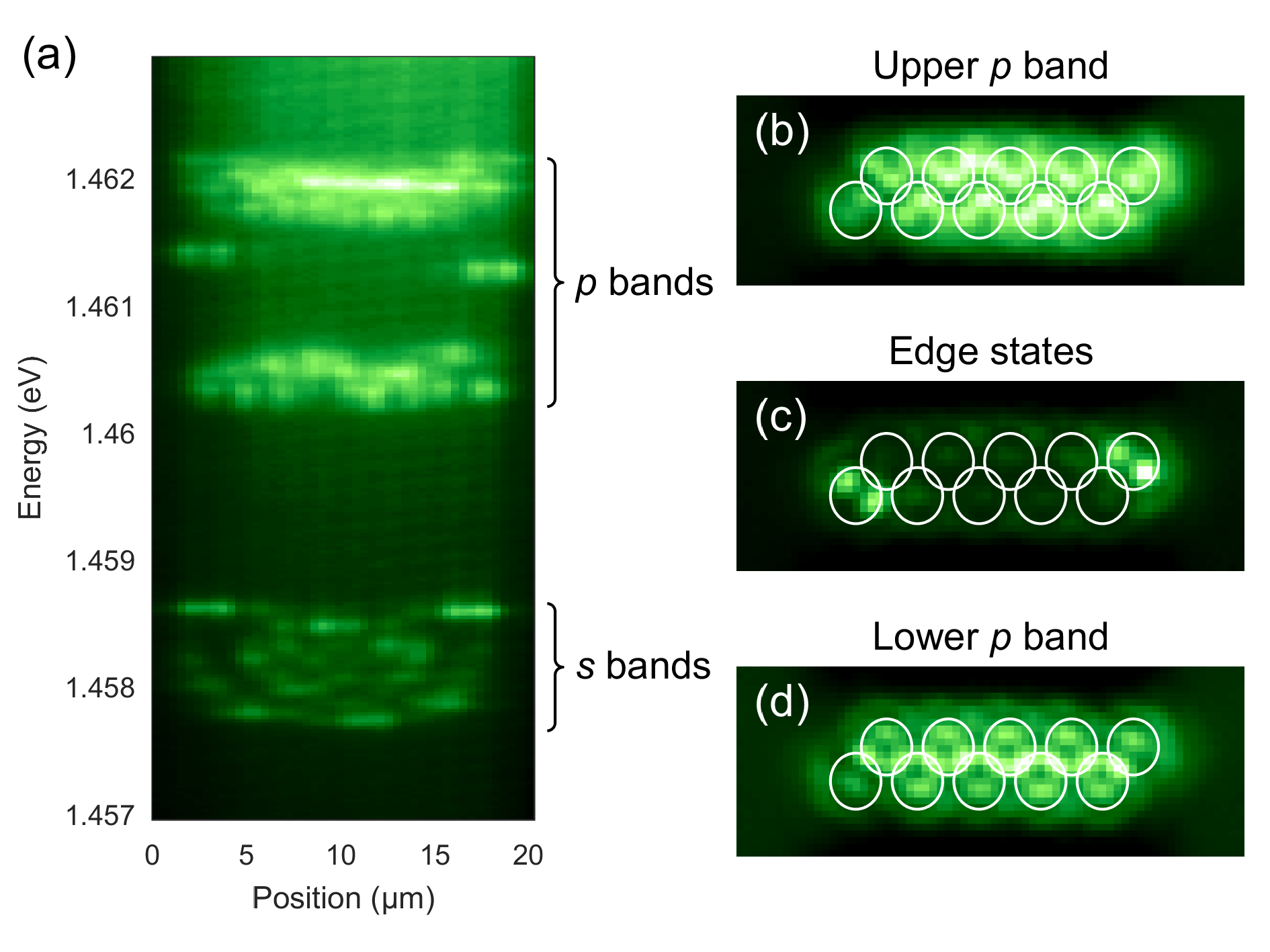}
\includegraphics[scale=1]{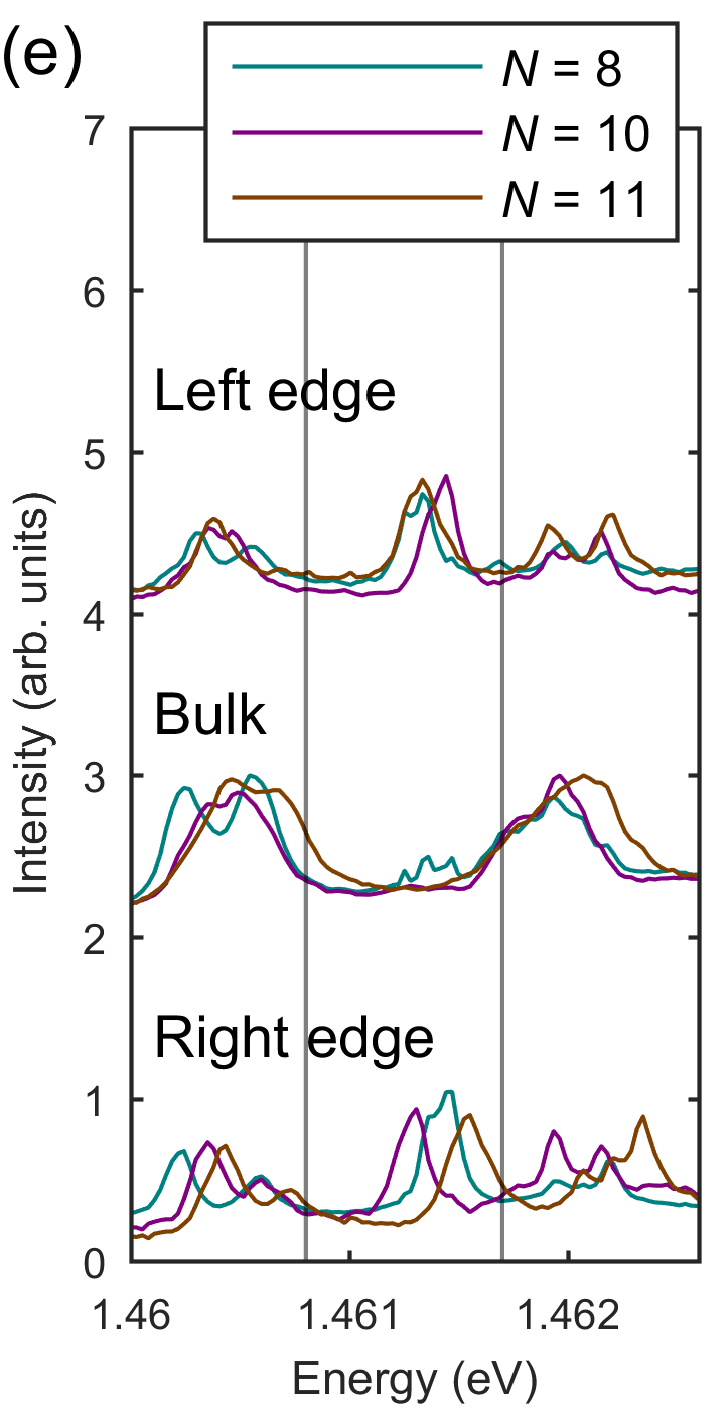}
\caption{(a) PL spectrum of a 10-site zigzag chain showing $s$ and $p$ bands. The real space images to the right show the emission integrated over the upper $p$ band (b), the edge states (c) and the lower $p$ bands (d). (e) Experimental spectra from the left edge, bulk and right edge of chains with 8, 10 and 11 sites.}
\label{zigzag_unpol}
\end{figure}

Considering our zigzag chain without the polarization degree of freedom we have a twofold SSH model \cite{St-Jean2017} which is shown in Figs. \ref{zigzag_unpol}(a)--(d) for the 10-site chain. In the real space spectrum of the chain in (a), we see $s$ and $p$ bands spanning approximately 1 and 2 meV respectively and separated by a gap of $\sim$1.5 meV. Due to the directionally-independent tunneling energy of $s$ orbitals, no gap opens between the bonding and anti-bonding bands. In the $p$ orbital case, the large difference in tunneling energies between the ``head-to-head'' and ``shoulder-to-shoulder'' types of bonding results in the opening of a large gap of about 1 meV. Modes existing within the gap are clearly seen at the edges of the chain. In Figs. \ref{zigzag_unpol}(b)--(d) we show the real space emission from the upper (anti-bonding) $p$ bands, the edge states and the lower (bonding) $p$ bands. In the bonding/anti-bonding bands, the constituent states are delocalized across the chain, as can be observed in the real space images, where the respective symmetries are visualized by clear intensity lobes/nodes across the junctions connecting pillars. For the modes within the gap, however, the intensity is predominantly localized on the outermost pillars of the chain, and the mode is in the $p_y$ subspace in both cases.
 
In addition to our 10-site chain, we also measured the real space PL spectra of chains with 8 and 11 sites. The results for all three chains are shown in Fig. \ref{zigzag_unpol}(e). In spectra from the bulk of the chains, corresponding to the whole chain excluding the outermost pillars, distinct lower and upper $p$ bands can be seen, separated by a gap delimited by vertical lines. In the spectra from the edges, corresponding to the outermost pillars, emission can be seen from states within the lower and upper $p$ bands, in addition to modes residing in the gap. 

\section{Spin-orbit coupled $P$ modes of a single pillar}

\begin{figure}[t]
\centering
\includegraphics[scale=0.9]{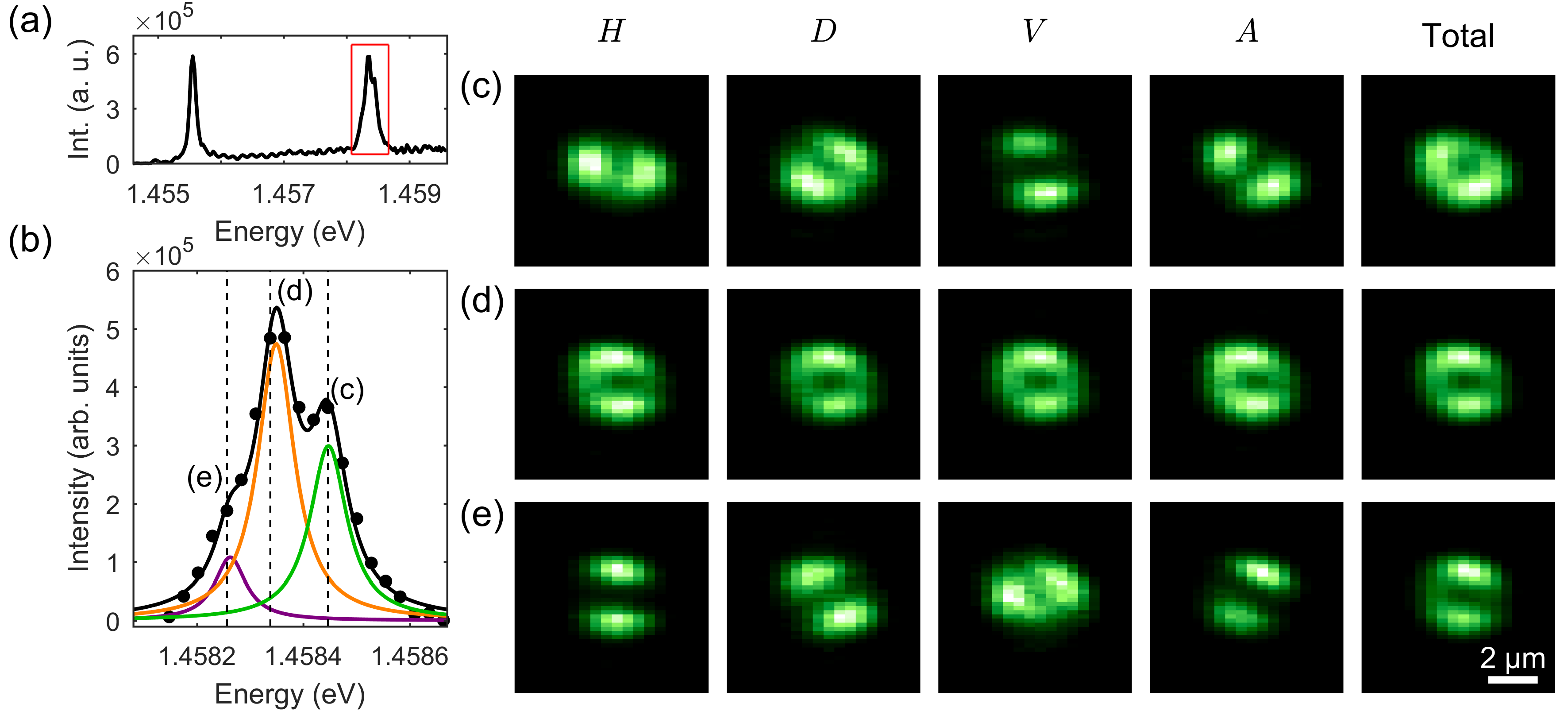}
\caption{(a) PL spectrum of a single pillar showing $s$ and $p$ mode peaks. (b) Close-up of the fine structure-split $p$ mode spectrum showing experimental data with three fitted underlying peaks. The real space images show emission resolved in horizontal ($H$), diagonal ($D$), vertical ($V$) and anti-diagonal ($A$) polarizations of the upper (c), middle (d) and lower (e) fine structure states, which correspond respectively to a radial spin vortex, a superposition of orthogonal hyperbolic spin anti-vortices and an azimuthal spin vortex \cite{PhysRevLett.115.246401}.}
\label{pillarmodes}
\end{figure}

In order to probe the eigenmodes of a single micropillar, we perform energy-resolved real space PL measurements on a single pillar resolved in horizontal ($H$), diagonal ($D$), vertical ($V$) and anti-diagonal ($A$) polarizations, shown in Fig. \ref{pillarmodes}. We find the strong TE-TM splitting in the sample splits the $p$ orbital into modes at three different energies, as can be seen in the spectrum shown in Fig. \ref{pillarmodes}(b). In the real space emission at these energies, the total intensities show ring-like profiles, similar to Laguerre-Gauss modes. In the upper and lower peaks the polarization points radially and azimuthally respectively as can be seen in the rotation of the modes with detection polarization. The middle peak on the other hand shows no rotation. A detailed analysis of this effect is provided in Ref. \cite{PhysRevLett.115.246401}.

\section{Polarization-resolved spectrum of two coupled pillars}

\begin{figure}[h]
\centering
\includegraphics[scale=1]{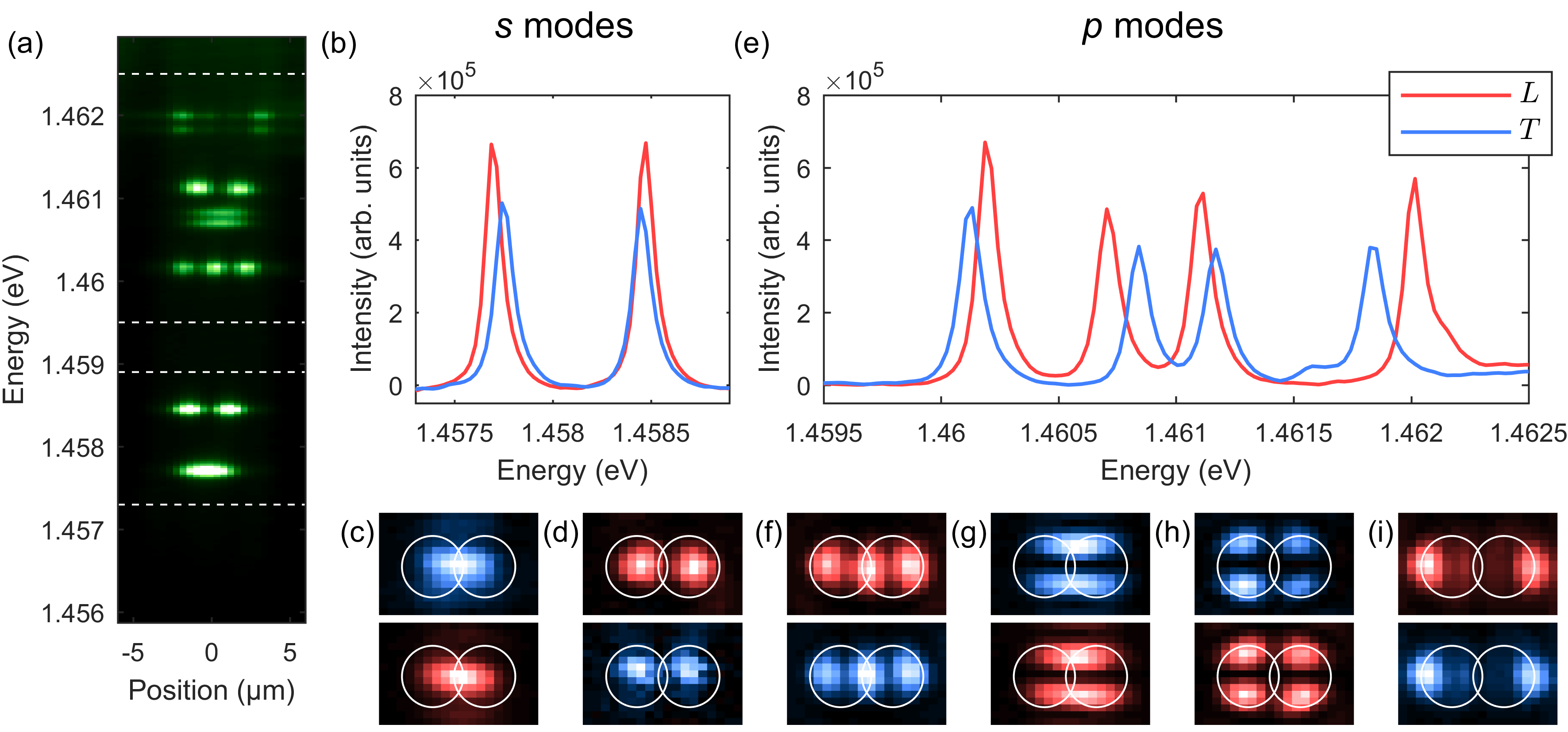}
\caption{Experimental single dimer energy spectrum. (a) Real space PL spectrum showing energy against longitudinal position of the dimer. (b) Polarization-resolved spectrum of the hybridized $s$ modes corresponding to the lower two dashed white lines in (a). Panels (c) and (d) show differential real space images of the bonding and anti-bonding $s$ modes respectively. (e) Polarization-resolved spectrum of the hybridized $p$ modes corresponding to the upper two dashed white lines in (a). Panels (f) and (i) show differential real space images of the bonding and anti-bonding $p_x$ modes respectively, and (g) and (h) show differential real space images of the bonding and anti-bonding $p_y$ modes respectively. The two detection polarizations are longitudinal ($L$) and transverse ($T$) to the tunneling direction. A baseline has been subtracted in (e).}
\label{dimer_pol}
\end{figure}

In order to estimate the magnitude of the different tunneling processes between the $p$ modes of overlapping micropillars we studied the simplest possible case of a single dimer (two 3 $\mu$m pillars separated by 2.55 $\mu$m). Measurements of the energy spectrum are presented in Fig. \ref{dimer_pol}. In Fig. \ref{dimer_pol}(a) we see an energy-resolved real space image along the long axis of the dimer. The set of two modes at lower energy arise from hopping between $s$ orbitals, and the higher energy modes are the hybridized $p$ modes. The twofold degeneracy of the $p$ modes doubles the number of bonded modes, and the different symmetries of the $p_x$ and $p_y$ orbitals leads to four modes rather than two. Physically, in a dimer oriented along $x$, $p_x$ orbitals have larger spatial overlap than $p_y$ orbitals, which is why their tunneling energy is so much larger. Even without resolving in polarization, it is clear from Fig. \ref{dimer_pol}(a) that there is some splitting of the four bonded modes. 

Fig. \ref{dimer_pol}(b) shows the polarization-resolved $s$ mode spectrum. The hopping energy is 0.37 meV, and there is a small but finite polarization splitting induced purely by polarization-dependent tunneling, since there is no on-site splitting term for modes with $l=0$. As expected, $H$ polarization, which is longitudinal to the tunneling link between the two pillars, has a larger tunneling energy than $V$ polarization, which is transverse. This is evidenced by the larger separation between the two $H$ peaks compared to the $V$ peaks. The difference in tunneling energies for the two polarizations is around 40 $\mu$eV, and corresponding differential real space images of the bonding and anti-bonding modes are shown in Figs. \ref{dimer_pol}(c) and (d). For the $p$ modes, we estimate a hopping energy of 0.88 meV between $p_x$ orbitals and 0.18 meV between $p_y$ orbitals, which is five times smaller. In this case, polarization corrections to both the on-site and hopping energies are present: we estimate the former has a magnitude of around 0.11 meV and the latter has values on the order of 70 $\mu$eV and 40 $\mu$eV for $p_x$ and $p_y$ orbitals respectively.

\section{Tight-binding model}

In order to develop our tight-binding model we first calculate the effect of the TE-TM splitting on the energy levels of a single pillar. To this end it is possible to use a degenerate perturbation theory starting from the Hamiltonian written in the basis of $H$ and $V$ polarized polaritons,

\begin{equation}
H_{\mathrm{pillar}}=
\begin{pmatrix}
\frac{1}{2m}(\hat{p}_x^2+\hat{p}_y^2)+U(x,y)+\beta (-\hat{p}_x^2+\hat{p}_y^2) & 2\beta \hat{p}_x \hat{p}_y \\
2\beta \hat{p}_x \hat{p}_y & \frac{1}{2m}(\hat{p}_x^2+\hat{p}_y^2)+U(x,y)-\beta (-\hat{p}_x^2+\hat{p}_y^2)  \\
\end{pmatrix},
\end{equation}

\noindent
where $m$ is the polariton mass, $\beta$ represents the strength of the TE-TM splitting, $\omega$ is the strength of the harmonic confinement $U(x,y)=\frac{1}{2}m\omega^2(x^2+y^2)$, and $\hbar=1$ \cite{PhysRevLett.115.246401}. This Hamiltonian can be written as the sum of two terms: a two-dimensional harmonic potential for each polarization component ($H$ and $V$), for which the eigenmodes are the $p$ orbitals described in the main paper, and a TE-TM perturbation term ($H_{\mathrm{pert}}$) that includes all the terms depending on $\beta$. To calculate the effect of TE-TM splitting on the $p$ orbitals one needs to calculate the integrals $\langle p_i^j|H_{\mathrm{pert}}|p_l^n \rangle$, where $i,l$ are either $x$ or $y$ and $j,n$ are either $H$ or $V$. These are a total of 16 integrals composing a $4\times 4$ matrix, but only 10 are independent due to the Hermitian nature of the Hamiltonian. Thus, taking into account the symmetries of this problem (i.e. the rotational symmetry of the harmonic potential and the $x$ and $y$ direction of the coupling between pillars), the effect of the perturbation can be written in a compact matrix form as:

\begin{equation}
H_0=\psi^{\dagger}\,E_s\,\psi=\psi^{\dagger}
\begin{pmatrix}
\Delta E& 0 & 0 & \Delta E\\
0 & -\Delta E& \Delta E& 0 \\
0 & \Delta E& -\Delta E& 0 \\
\Delta E& 0 & 0 & \Delta E\\ 
\end{pmatrix}
\psi,
\end{equation}

\noindent
with $\psi^{\dagger}=(p_x^{H},p_y^{H},p_x^{V},p_y^{V})$, and $\Delta E$ being the energy shift induced by the spin-orbit coupling. Here, the matrix $E_s$ describes the perturbation, induced by the TE-TM splitting, to the energy eigenmodes of the first excited manifold of the two-dimensional homogenous harmonic oscillator. This is exactly the same result as in Ref. \cite{PhysRevLett.115.246401}, but written in the $p$ orbital basis instead of the spin-vortex one. Secondly, we want to calculate the hopping terms in the presence of the TE-TM splitting. To this end we start from the Hamiltonian of a dimer (also written in the basis of $H$ and $V$ polarized polaritons),

\begin{equation}
H_{\mathrm{dimer}}=
\begin{pmatrix}
(p_x^2+p_y^2)/2m+U_{d}(x,y)+\beta (-p_x^2+p_y^2) & 2\beta p_x p_y \\
2\beta p_x p_y & (p_x^2+p_y^2)/2m+U_{d}(x,y)-\beta (-p_x^2+p_y^2)  \\
\end{pmatrix},
\end{equation}

\noindent
where the term $U_{d}(x,y)=min[U(x-x_0,y-y_0),U(x+x_0,y+y_0)]$ describes two harmonic potentials centered at $+(x_0,y_0)$ and $-(x_0,y_0)$. To calculate the hopping terms in the presence of the TE-TM splitting, we use again the degenerate perturbation theory and re-write the Hamiltonian as $H=H_{-}+H_{+}+H_{\mathrm{pert}}'$, with

\begin{equation}
H_{-}=
\begin{pmatrix}
(p_x^2+p_y^2)/2m+U(x-x_0,y-y_0) & 0 \\
0 & (p_x^2+p_y^2)/2m+U(x-x_0,y-y_0)  \\
\end{pmatrix},
\end{equation}

\begin{equation}
H_{+}=
\begin{pmatrix}
(p_x^2+p_y^2)/2m+U(x+x_0,y+y_0) & 0 \\
0 & (p_x^2+p_y^2)/2m+U(x+x_0,y+y_0)  \\
\end{pmatrix},
\end{equation}

\noindent
and

\begin{equation}
H_{\mathrm{pert}}'=H_{\mathrm{pert}}+\mathbb{I}\Delta U(x,y)=
\begin{pmatrix}
\Delta U(x,y)+\beta (-p_x^2+p_y^2) & 2\beta p_x p_y \\
2\beta p_x p_y & \Delta U(x,y)-\beta (-p_x^2+p_y^2)  \\
\end{pmatrix},
\end{equation}

\noindent
where $\Delta U(x,y)=U_{d}(x,y)-U(x-x_0,y-y_0)-U(x+x_0,y+y_0)$. In order to evaluate the hopping terms one needs again to evaluate the integrals $\langle p_i^j|H_{\mathrm{pert}}'|p_l^n \rangle$. This time, however, the eigenmodes to use are the eigenmodes of $H_{+}$ and $H_{-}$, which are the $p_x$ and $p_y$ orbitals in the $H$ and $V$ polarization centered at $\pm(x_0,y_0)$. Moreover, in this case $H_{\mathrm{pert}}'$, together with the terms depending on $\beta$, also has additional terms depending on $\Delta U(x,y)$ describing the potential barriers between the two pillars. Note that even if $\Delta U(x,y)$ goes to $-\infty$ with $x,y\rightarrow \pm\infty$, the perturbation approach still gives meaningful results because the $p$ orbitals go to zero much faster than the potential ($\propto e^{-(x^2+y^2)}$ rather than $\propto (x^2+y^2)$), and therefore the contribution to the integral at large radii is negligible. To evaluate the hopping terms, the integrals have to be evaluated for the two cases of hopping along the $x$ and $y$ directions. This has two main consequences, first that two different masses have to be used in the confining potential for the propagation of a $H$ or $V$ polarized mode along two different directions, second that one has to evaluate a total of 20 integrals, 10 for each direction. Amongst these, it turns out that, for symmetry reasons, 10 are identically zero (5 in each direction) and that the 5 remaining non-zero integrals relative to one propagation direction have an identical counterpart integral relative to the other propagating direction. Therefore there is a total of 5 different hopping terms, which are given by:

\begin{eqnarray}
t_a^{\parallel}&=&\frac{\omega e^{-a^2m_{\parallel}\omega}}{2\sqrt{\pi}}
\left[4a\sqrt{m_{\parallel}\omega}(-1+a^2m_{\parallel}\omega)+\sqrt{\pi}(-2+m_{\parallel}(-2\beta+a^2\omega(1+10m_{\parallel}\beta)+2a^4m_{\parallel}\omega^2(1-2m_{\parallel}\beta)))\right]
\nonumber
\\
t_a^{\perp}&=&\frac{\omega e^{-a^2m_{\perp}\omega}}{2\sqrt{\pi}}
\left[4a\sqrt{m_{\perp}\omega}(-1+a^2m_{\perp}\omega)+\sqrt{\pi}(-2+m_{\perp}(2\beta+a^2\omega(1-10m_{\perp}\beta)+2a^4m_{\perp}\omega^2(1-2m_{\perp}\beta)))\right]
\nonumber
\\
t_t^{\parallel}&=&\frac{\omega e^{-a^2m_{\parallel}\omega}}{2\sqrt{\pi}}
\left[-2a\sqrt{m_{\parallel}\omega}+\sqrt{\pi}(-2+2m_{\parallel}\beta+a^2m_{\parallel}\omega(-1+2m_{\parallel}\beta))\right]
\nonumber
\\
t_t^{\perp}&=&\frac{\omega e^{-a^2m_{\perp}\omega}}{2\sqrt{\pi}}
\left[-2a\sqrt{m_{\perp}\omega}-\sqrt{\pi}(2+2m_{\perp}\beta+a^2m_{\perp}\omega(1+2m_{\perp}\beta))\right]
\nonumber
\\
t_m&=&8\omega\beta m_{\parallel}^2 m_{\perp}^2 e^{-\frac{2a^2m_{\parallel}m_{\perp}\omega}{m_{\parallel}+m_{\perp}}}
\frac{[-m_{\perp}+m_{\parallel}(-1+4a^2m_{\perp}\omega)]}{(m_{\parallel}+m_{\perp})^4}.
\nonumber
\end{eqnarray}

\noindent
Here, $m_{\parallel}$ and $m_{\perp}$ represent the masses of modes propagating with polarization parallel and perpendicular, respectively, to the propagation direction, and $a$ is half the center-to-center distance between pillars. The term $\tau_a^{\parallel}$ ($\tau_a^{\perp}$) describes the hopping amplitude of an orbital aligned along the propagation direction with polarization parallel (perpendicular) to it, i.e. either $p_x^H$ ($p_x^V$) propagating along the $x$ direction or $p_y^V$ ($p_y^H$) propagating along the $y$ direction. Similarly, $\tau_t^{\parallel}$ and $\tau_t^{\perp}$ describe the hopping amplitudes of orbitals with alignment perpendicular to the propagation direction. Finally, $\tau_m$ describes the hopping amplitude for a given orbital that hybridizes while propagating. These terms can be written in a compact matrix form as:

\begin{equation*}
H_{\tau_x}=\psi^{\dagger}
\begin{pmatrix}
\tau_a^{\parallel} & 0 & 0 & 0 \\
0 & \tau_{t}^{\parallel} & 0 & 0 \\
0 & 0 & \tau_{a}^{\perp} & 0 \\
0 & 0 & 0 & \tau_{t}^{\perp} \\ 
\end{pmatrix}
\psi,\,\,\,\,
H_{\tau_y}=\psi^{\dagger}
\begin{pmatrix}
\tau_t^{\perp} & 0 & 0 & 0 \\
0 & \tau_{a}^{\perp} & 0 & 0 \\
0 & 0 & \tau_{t}^{\parallel} & 0 \\
0 & 0 & 0 & \tau_{a}^{\parallel} \\ 
\end{pmatrix}
\psi,\,\,\,\,
H_{\tau_m}=\psi^{\dagger}
\begin{pmatrix}
0 & 0 & 0 & \tau_m \\
0 & 0 & \tau_m & 0 \\
0 & \tau_m & 0 & 0 \\
\tau_m & 0 & 0 & 0 \\ 
\end{pmatrix}
\psi.
\end{equation*}

\begin{figure}[h]
\centering
\includegraphics[scale = 1]{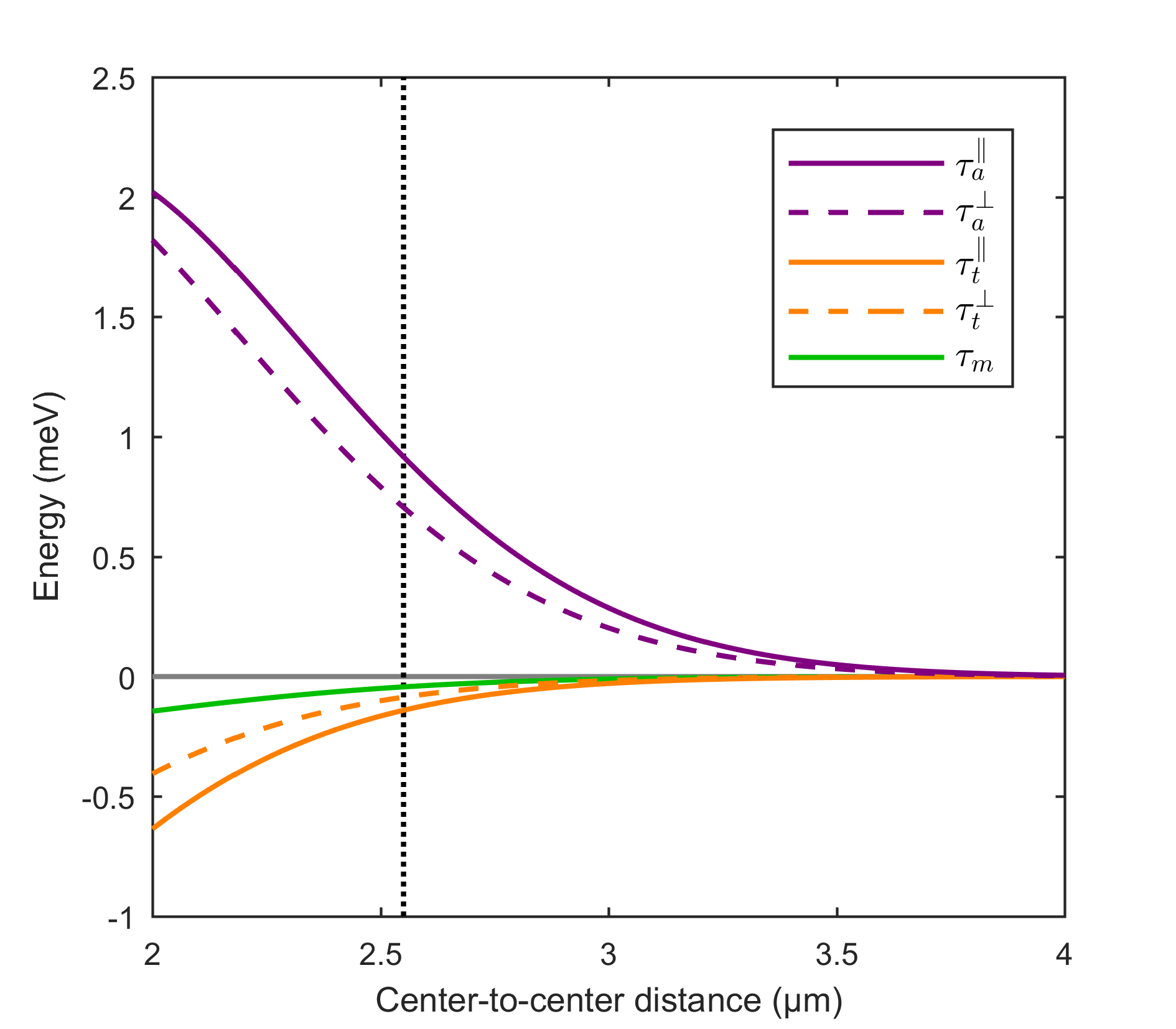}
	\caption{Hopping terms as a function of the distance between two pillar centers. Here the parameters are: $\hbar\omega=2.8$ meV, $\beta=-0.186$ meV$\cdot\mu$m$^2$, and $m_{\parallel}=8.08\times 10^{-5} m_0$ and $m_{\perp}=7.70\times 10^{-5} m_0$ ($m_0$ being the bare electron mass). The vertical line is set at 2.55 $\mu$m corresponding to the experimental pillar distance.}
\label{hterms}
\end{figure}

\noindent
Figure \ref{hterms} shows the strength of the hopping terms, as a function of center-to-center distance, for given values of $\omega,\beta$ and $m$. Here the parameter $\hbar\omega$ is obtained from the experimental separation between $s$ and $p$ modes on a single pillar, $\beta$ from the effective mass difference in a nearby planar section of the cavity, and $m_{\parallel}$ and $m_{\perp}$ are used as fitting parameters. At a center-to-center distance of $2.55$ $\mu$m, the two $\tau_a$ terms have the highest absolute values ($0.91$ and $0.81$ meV) and are positive, the two $\tau_t$ terms are the second highest and have opposite sign with respect to the previous ones ($-0.14$ and $-0.11$ meV), and $\tau_m$ is the weakest one ($\approx -0.03$ meV) and barely visible on this energy scale. These values for the hopping parameters are in very good agreement with the experimental ones. Note, however, that the values of the masses used here is about 4 times bigger than the experimental ones. This difference can be understood observing that at a distance of $2.55$ $\mu$m two pillars already have a significant overlapping region, and therefore the perturbation approach only constitutes a rough approximation of the physical system. In addition, the confinement potential has a rectangular profile rather than the parabolic one used here. 

\section{Topological properties of the system}

In order to study the effect of the TE-TM splitting on the topology of a fourfold SSH polariton chain we start revising the topology of the simple SSH model. The bulk Hamiltonian in $k$ space for this model can be written as \cite{asboth2016short}:

\begin{equation}
H_{\rm{SSH}}(k)=
\begin{pmatrix}
0 & \tau_{\rm{intra}}+\tau_{\rm{inter}}e^{-i k a}\\
\tau_{\rm{intra}}+\tau_{\rm{inter}}e^{+i k a} & 0 \\
\end{pmatrix},
\end{equation}

\noindent
where $\tau_{\rm{intra}}$ and $\tau_{\rm{inter}}$ are the intra- and inter-dimer hopping. This Hamiltonian displays a time-reversal symmetry
\begin{equation} \mathbb{T}H(k)\mathbb{T}^{-1} =H(-k)
\end{equation} 
with $\mathbb{T}=\mathbb{I}\kappa$, where $\kappa$ is the complex conjugation operator,
as well as a chiral symmetry
\begin{equation}
\Gamma H(k)\Gamma^{-1} =-H(k)
\end{equation} 
with $\Gamma=\sigma_z$ a 2$\times$2 Pauli matrix, and
a charge-conjugation symmetry
\begin{equation}
 \mathbb{C}H(k)\mathbb{C}^{-1} =-H(-k)
\end{equation} 
with $\mathbb{C}=\sigma_z\kappa$. All these operators square to $1$ and the SSH model can be identified to be a $\mathbb{Z}$ topological insulator belonging to the BDI symmetry class \cite{1367-2630-12-6-065010}. To study the fourfold case we first derive the bulk Hamiltonian:

\begin{equation}
H^p_{\rm{SSH}}(k)=
\begin{pmatrix}
E_s & \tau(k) \\
\tau(k)^{\dagger} & E_s\\
\end{pmatrix}
\end{equation}

\noindent
with:

\begin{equation}
E_s=
\begin{pmatrix}
\Delta E& 0 & 0 & \Delta E\\
0 & -\Delta E& \Delta E& 0 \\
0 & \Delta E& -\Delta E& 0 \\
\Delta E& 0 & 0 & \Delta E\\
\end{pmatrix}
\hspace{0.5cm}
\tau(k)=
\begin{pmatrix}
\tau_a^{\parallel}+\tau_{t}^{\perp}e^{-i k a} & 0 & 0 & \tau_m(1+e^{-i k a}) \\
0 & \tau_t^{\parallel}+\tau_{a}^{\perp}e^{-i k a} & \tau_m(1+e^{-i k a}) & 0 \\
0 &  \tau_m(1+e^{-i k a}) & \tau_a^{\perp}+\tau_{t}^{\parallel}e^{-i k a} & 0 \\
\tau_m(1+e^{-i k a}) & 0 & 0 & \tau_t^{\perp}+\tau_{t}^{\parallel}e^{-i k a} \\
\end{pmatrix}.
\end{equation}

\noindent
Here, as before, $E_s$ represents the perturbation induced by the TE-TM splitting on the energies of the eigenmodes of each pillar, and $\tau(k)$ is the hopping between the two pillars in the unit cell, to which we refer as A and B. The simple case with $\Delta E=0$, $\tau_m=0$, and $\tau_a^{\parallel}=\tau_a^{\perp}$ and $\tau_t^{\parallel}=\tau_t^{\perp}$, i.e. without TE-TM splitting, still represents a $\mathbb{Z}$ topological insulator since it merely represents four 2$\times$2 SSH models \cite{St-Jean2017}.

Next we study the case when only the on-site TE-TM splitting is different from zero (i.e. $\Delta E\ne 0$ but $\tau_m=0$, and $\tau_a^{\parallel}=\tau_a^{\perp}$ and $\tau_t^{\parallel}=\tau_t^{\perp}$). To study the topological properties of this system we start looking for an operator implementing the chiral symmetry:

\begin{equation}
\Gamma H^p_{\rm{SSH}}(k) \Gamma^{-1} =-H^p_{\rm{SSH}}(k),
\end{equation}

\noindent
and observe that the transforming operators will be the product of two-dimensional matrices acting on the A-B sublattice with four-dimensional matrices acting in the mode space of the individual pillars. The matrices acting on the A-B sublattice are the three Pauli matrices and their effect on the diagonal block matrices ($E_s$) is none, since the two blocks are identical. Therefore the effect of the transformation on the four-dimensional pillar-mode subspace has to be to introduce a minus sign. There are only four 4$\times$4 matrices that satisfy this condition:

\begin{equation}
\gamma_1=
\begin{pmatrix}
0 & 1 & 0 & 0 \\
1 & 0 & 0 & 0 \\
0 & 0 & 0 & -1 \\
0 & 0 & -1 & 0 \\
\end{pmatrix}
\hspace{1cm}
\gamma_2=
\begin{pmatrix}
0 & -i & 0 & 0 \\
i & 0 & 0 & 0 \\
0 & 0 & 0 & -i \\
0 & 0 & i & 0 \\
\end{pmatrix}
\end{equation}

\begin{equation}
\gamma_3=
\begin{pmatrix}
0 & 0 & 1 & 0 \\
0 & 0 & 0 & 1 \\
-1 & 0 & 0 & 0 \\
0 & -1 & 0 & 0 \\
\end{pmatrix}
\hspace{1cm}
\gamma_4=
\begin{pmatrix}
0 & 0 & 1 & 0 \\
0 & 0 & 0 & -1 \\
1 & 0 & 0 & 0 \\
0 & -1 & 0 & 0 \\
\end{pmatrix}.
\end{equation}

\noindent
The effect of the two-dimensional Pauli matrices on the off-diagonal hopping matrices $\tau(k)$ and $\tau^{\dagger}(k)$, is either to add a minus sign, or to swap them, or to swap them and add a minus sign. In the first case this means that at least one of the four matrices $\gamma_i$, $i=1,2,3,4$ have to implement the transformation $\gamma_i\tau(k)\gamma_i^{-1}=\tau(k)$ and $\gamma_i\tau(k)^{\dagger}\gamma_i^{-1}=\tau(k)^{\dagger}$. In the second and third case one of the four matrices has to implement either the transforms: $\gamma_i\tau(k)\gamma_i^{-1}=\tau(k)^{\dagger}$ and $\gamma_i\tau(k)^{\dagger}\gamma_i^{-1}=\tau(k)$, or $\gamma_i\tau(k)\gamma_i^{-1}=-\tau(k)^{\dagger}$ and $\gamma_i\tau(k)^{\dagger}\gamma_i^{-1}=-\tau(k)$. It is easy to check that both $\gamma_3$ and $\gamma_4$ implement the transformations: $\gamma_i\tau(k)\gamma_i^{-1}=\tau(k)$ and $\gamma_i\tau(k)^{\dagger}\gamma_i^{-1}=\tau(k)^{\dagger}$. Therefore, $\Gamma$ can be either $\Gamma_1=\sigma_z\otimes \gamma_3$ or $\Gamma_2=\sigma_z\otimes \gamma_4$, in the first case $\Gamma^2=1$, in the second $\Gamma^2=-1$. Then we look for operators implementing the time reversal and charge conjugation symmetries. For the time reversal symmetry one can proceed in a similar way as for the chiral symmetry. Since the Pauli matrices acting on the A-B space have no effect on the diagonal block matrices, since the two blocks are identical, the effect of the transformation on the internal four-dimensional space has to be to leave it unchanged, i.e. $t_iE_st_i^{-1}=E_s$. There are, again, four matrices that satisfy this condition:

\begin{equation}
t_1=
\begin{pmatrix}
1 & 0 & 0 & 0 \\
0 & 1 & 0 & 0 \\
0 & 0 & 1 & 0 \\
0 & 0 & 0 & 1 \\
\end{pmatrix}
\hspace{1cm}
t_2=
\begin{pmatrix}
1 & 0 & 0 & 0 \\
0 & -1 & 0 & 0 \\
0 & 0 & -1 & 0 \\
0 & 0 & 0 & 1 \\
\end{pmatrix}
\end{equation}

\begin{equation}
t_3=
\begin{pmatrix}
0 & 0 & 0 & 1 \\
0 & 0 & 1 & 0 \\
0 & 1 & 0 & 0 \\
1 & 0 & 0 & 0 \\
\end{pmatrix}
\hspace{1cm}
t_4=
\begin{pmatrix}
0 & 0 & 0 & -i \\
0 & 0 & i & 0 \\
0 & i & 0 & 0 \\
-i & 0 & 0 & 0 \\
\end{pmatrix}.
\end{equation}

\noindent
As before, one now needs to study the effect of the Pauli matrices on the off-diagonal terms ($\tau(k)$, and $\tau^{\dagger}(k)$). Both $\sigma_y$ and $\sigma_z$ introduce a minus sign to $\tau(k)$ and $\tau^{\dagger}(k)$ while swapping them or not. But none of the $t_i$ matrices above introduces a minus sign to compensate for it. Moreover, $\sigma_x$ does not introduce a minus sign but swaps $\tau(k)$ and $\tau^{\dagger}(k)$, which is equivalent to changing $k\rightarrow -k$. This change in sign for $k$, however, is also introduced by the complex conjugation operator $\kappa$, and therefore also $\sigma_x$ has to be discarded. The only possible choice remaining for the A-B sublattice is the identity matrix. It is easy to check that both $\mathbb{T}_1=\mathbb{I}t_1\kappa$ and $\mathbb{T}_2=\mathbb{I}t_2\kappa$ implement the time reversal symmetry. In both cases $\mathbb{T}^2=1$. Therefore the system is still in the BDI symmetry class and is a topological $\mathbb{Z}$ insulator. It is useful to note here that $\mathbb{T}_1\mathbb{T}_2=\mathbb{I}t_2=\Gamma_1\Gamma_2$, and that $\mathbb{I}t_2$ is a symmetry of our Hamiltonian $H_{\rm{SSH}}(k)$ since $(\mathbb{I}t_2)H_{\rm{SSH}}(k)(\mathbb{I}t_2)^{-1}=H_{\rm{SSH}}(k)$.

Finally we want to consider the case when also the effect of the TE-TM splitting on the propagation of particles along the chain is non-zero (i.e. $\Delta E\ne 0$ but $\tau_m\ne 0$, and $\tau_a^{\parallel}\ne\tau_a^{\perp}$ and $\tau_t^{\parallel}\ne\tau_t^{\perp}$). Starting from the chiral symmetry, it is easy to check that none of the required identities: $\gamma_i\tau(k)\gamma_i^{-1}=\tau(k)$ and $\gamma_i\tau(k)^{\dagger}\gamma_i^{-1}=\tau(k)^{\dagger}$, or $\gamma_i\tau(k)\gamma_i^{-1}=\tau(k)^{\dagger}$ and $\gamma_i\tau(k)^{\dagger}\gamma_i^{-1}=\tau(k)$, or $\gamma_i\tau(k)\gamma_i^{-1}=-\tau(k)^{\dagger}$ and $\gamma_i\tau(k)^{\dagger}\gamma_i^{-1}=-\tau(k)$, are satisfied by any of the four $\gamma_i$, and that, therefore, the chiral symmetry is broken. Since, instead, the time reversal symmetry is not broken (both $\mathbb{T}_1$ and $\mathbb{T}_2$ defined before are still implementing the time reversal symmetry) the system, in this case, is not a topological insulator.

\section{Control of the TE-TM splitting through layer structure design}

As mentioned in the main text, the two major contributions to the polarization splitting in our micropillar array come from the layer structure of the microcavity heterostructure itself, and also due to in-plane confinement after etching. In principle, one may control the first contribution through layer structure design (see Ref. \cite{PhysRevB.59.5082} for a detailed theoretical treatment) by varying the center of the reflectivity stopband, through the precise layer structure in the Bragg mirrors, while keeping the planar cavity mode fixed. Control over the TE-TM splitting contribution from the mirrors means the TE-TM $\beta$ factor can take arbitrary (positive or negative) values. This means the \em{}on-site\rm{} polarization splitting of the $p$ modes in micropillars can be engineered through epitaxial growth to be very large or vanishingly small. 

As for the polarization splitting in the \em{hopping}\rm{} energy of modes, it is determined by the different barrier heights between pillars imposed by (and determined by the width of) the junction where they overlap. In our case the pillars are circular, but other single and coupled pillar geometries \cite{PhysRevLett.81.2582} may offer advantages and allow precise tailoring of the hopping polarization dependence.

\end{document}